\documentclass{article}

\usepackage[LGR,T1]{fontenc} 
\usepackage[utf8]{inputenc} %
\usepackage{substitutefont} 
\usepackage[polutonikogreek,english]{babel}
\usepackage[largesc]{newtxtext} %
\usepackage[varqu,varl]{zi4}
\usepackage{cabin}
\usepackage[vvarbb,upint,slantedGreek,smallerops]{newtxmath}
\substitutefont{LGR}{\rmdefault}{Tempora} 

\usepackage{textcomp}

\usepackage[margin=1.5in]{geometry}
\usepackage{hyperref}
\usepackage{url}
\usepackage{amsmath}
\usepackage{amsfonts}
\usepackage{amssymb}

\usepackage[pdftex]{graphicx}
\usepackage{lastpage}
\usepackage{slashed}

\usepackage{color}

\usepackage{xfrac}

\usepackage{comment}
\usepackage[yyyymmdd,hhmmss]{datetime} 

\usepackage{lastpage}
\usepackage[running,mathlines]{lineno}
\usepackage{isomath}
\usepackage{subcaption}
\usepackage{siunitx}
\usepackage{upgreek}

\usepackage{marginnote}

\usepackage{marginnote}

\usepackage{amsthm}

\newtheorem{theorem}{Theorem}

\newtheorem{lemma}{Lemma}

\newtheorem{postulate}{Postulate}

\newtheorem*{law*}{Law}

\theoremstyle{definition}

\usepackage{verbatim}

\usepackage{xspace}
\usepackage{braket}
\usepackage{mathtools}
\mathtoolsset{showonlyrefs=true}

\usepackage{siunitx}
\usepackage{physics}

\newcommand{\pr}{\ensuremath{\mathrm{P}}}

\newcommand{\piu}%
{\textrm{\greektext p}}
\newcommand{\eu}%
{\ensuremath{\mathrm{e}}}
\newcommand{\iu}%
{\ensuremath{\mathrm{i}}}

\providecommand{\newoperator}[3]{%
\newcommand*{#1}{\mathop{#2}#3}}

\newcommand{\tran}%
{\textsf{T}}
\newcommand{\herm}%
{\textsf{H}}
\newcommand{\deltau}%
{\textrm{\greektext d}}

\makeatletter
\providecommand*{\diff}%
{\@ifnextchar^{\DIfF}{\DIfF^{}}}
\def\DIfF^#1{%
\mathop{\mathrm{\mathstrut d}}
\nolimits^{#1}\gobblespace}
\def\gobblespace{%
\futurelet\diffarg\opspace}
\def\opspace{%
\let\DiffSpace\!%
\ifx\diffarg(%
\let\DiffSpace\relax
\else
\ifx\diffarg[%
\let\DiffSpace\relax
\else
\ifx\diffarg\{%
\let\DiffSpace\relax
\fi\fi\fi\DiffSpace}

\newoperator{\loglike}%
{\mathrm{loglike}}{\nolimits}

\newoperator{\notloglike}%
{\mathrm{notloglike}}{\limits}

\makeatother

\usepackage{subcaption}

\newcommand{\Schroedinger}{Schr\"{o}dinger\xspace}

\title{A Theory for the Time Arrow}
\author{{Davi Geiger} and {Zvi M.\ Kedem}
 \\
 Courant Institute of Mathematical Sciences\\
 New York University}
\date{}

\newcommand{\normalx}[3]{\ensuremath{\mathcal{N}\left(#1 \mid #2,#3\right)}\xspace}

\newcommand{\thus}{\ensuremath{\quad \text{implying} \quad}\xspace}

 \newcommand{\tuu}{\ensuremath{\mathrm{tu}}\xspace}
 \newcommand{\duu}{\ensuremath{\mathrm{du}}\xspace}
 
 \newcommand{\entropy}{\ensuremath{\mathrm{entropy}}\xspace}
 \newcommand{\boson}{\ensuremath{\mathrm{boson}}\xspace}
 \newcommand{\fermion}{\ensuremath{\mathrm{fermion}}\xspace}
 \newcommand{\bos}{\ensuremath{\mathrm{b}}\xspace}
 \newcommand{\fer}{\ensuremath{\mathrm{f}}\xspace}
 \newcommand{\gr}{\ensuremath{\mathrm{g}}\xspace}
 \newcommand{\ph}{\ensuremath{\mathrm{p}}\xspace}

\renewcommand{\piu}{\uppi}
\renewcommand{\deltau}{\updelta}
\newcommand{\reals}{\ensuremath{\mathbb{R}}\xspace}

\let\originalpartial\partial
\let\partial\relax
\newrobustcmd*{\partial}{\text{\rotatebox[origin=t]{10}{\scalebox{0.95}[1]{\ensuremath{\originalpartial}}}}\hspace{-0.05em}}

\begin{document}

\maketitle
\thispagestyle{plain}
\begin{abstract}
\noindent
Physical laws for elementary particles can be described by the quantum dynamics equation
\[\mathrm{i} \hbar {\frac {\partial }{\partial t}} |\Psi_t\rangle=H |\Psi_t\rangle\;\quad \text{implying} \quad \; |\Psi_t\rangle=\mathrm{e}^{-\mathrm{i} \frac{1}{\hbar} \int_0^t H(t') \mathrm{d} t' } |\Psi_t\rangle\,,\]
where $|\Psi_t\rangle$ is the quantum state of a particle in Hilbert space and $H$ is the Hamiltonian.

A probability density function over $\mathbf{r}$ at time $t$ is given by $\rho (\mathbf{r},t)=|\langle\mathbf{r}|\Psi_t\rangle|^2$,
where $|\mathbf{r}\rangle$ is a position basis representation. Entropy  can be associated with these probability densities, characterizing the position information of a particle. Coherent states are localized wave packets and may describe the spatial distribution for some particle states. We show that due to a dispersion property of Hamiltonians in quantum physics, the entropy of coherent states increases over time. We investigate a partition of the Hilbert space into four sets based on whether the entropy is (i) increasing but not a constant, (ii) decreasing but not a constant, (iii) a constant, (iv) oscillating.

We propose a law that entropy (weakly) increases in time in quantum physics. Thus, states in set (ii) are disallowed, and the states in set (iii) cannot complete an oscillation period. 

Then, according to this law, quantum theory is not time-reversible unless the state is in set (iii), i.e., it is a stationary state (an eigenstate of the Hamiltonian.). This additional law in quantum theory limits physical scenarios beyond conservation laws, providing causality reasoning by defining an arrow of time.
\end{abstract}

\tableofcontents

\section{Introduction}

\subsection{Overview}

Much of the behavior of a physical system follows conservation laws, obtained by applying Noether's theorem \cite{noether1918invariante} to the symmetry transformations of the Lagrangian that models the system. However, conservation laws do not account for an arrow of time and therefore they cannot account for causality. Moreover,  both classical and quantum physics laws today are time-reversible.

A time arrow appears in physics only when statistics of multiple particles are introduced, as one derives the entropy function from the distribution of microstates, that is, microscopic states described by the positions and the momenta of the particles. Entropy is a measure of the number of possible microstates of a system, consistent with the thermodynamic properties of the macrostate. The second law of thermodynamics\cite{clausius1867mechanical} postulates that the entropy increases over time, technically increasing in a weakly manner, so it could be constant for any time interval.

Since such a law does not apply to individual particles, and there is no law to provide an arrow of time for a one-particle system, one cannot satisfactorily account for causality events. For example, how to answer the question: ``What causes an excited electron in the hydrogen atom to jump to the ground state while emitting radiation?'' While transition probabilities obtained from Fermi's golden rule \cite{dirac1927quantum,fermi1950nuclear} are highly accurate, this rule cannot be a causality explanation. Otherwise an energy perturbation method would be a source for the arrow of time. Similarly, we can ask: ``Why do nuclear decays occur?'', and again, despite accurate predictions, we do not have a causality explanation. Consider also the question: ``Why do high-speed particle colliding with each other transform into new particles?'' Such is the outcome of an electron-positron collision leading to two photons output.  If none of the events described above happened, conservation laws would still be satisfied. While conservation laws must be respected, there is no accounting of causality.

In order to address the question of causality we first review the entropy concept in quantum physics. Quantum theory introduces probability as intrinsic to the description of a one-particle system. A probability $\pr(o)$ is assigned to a specific value of $o \in \mathcal{O}$, where $ \mathcal{O}$ is an observable such as position, energy, or momentum. The observables can form a discrete set $\mathcal{O}=\{o_i;\, i = 1, 2, \dots, N\}$, where  $N$ can be arbitrarily large, or a continuous set $\mathcal{O}=\{o(\alpha)\mid \alpha \in \reals\}$. Such probability can then be associated with a measure of information about $ \mathcal{O}$. The more concentrated is the probability around a few given values of $\mathcal{O}$, the more information is provided about the observable. For a finite set $\mathcal{O}$ with $N$ possible outputs, the Shannon entropy, $\text{H}=-\sum_{i=1}^N \pr(o_i) \log_2 \pr(o_i)$, is a measure of such information. The larger is the entropy, the lesser is the information about the outcome of an observation of $\mathcal{O}$.

Extending the concept of entropy to continuous variables, continuous distributions, and to quantum physics has been challenging. For example, von Neumann's entropy \cite{von2018mathematical} requires the existence of classical statistics elements (mixed states) in order not to be zero, and consequently it assigns zero entropy to all one-particle systems. Our goal is to assign an entropy measure for a one-particle system that can be extended to multiple particles. Therefore, we do not consider von Newman entropy as a principal entropy measure. Attempts to take the limit of the discrete Shannon entropy as the number of output states goes to infinity and the interval between them goes to zero require the introduction of the distribution of the discretization lattice itself, leading to  inifinite constants  (see Jaynes \cite{jaynes1965gibbs}). Other entropy definitions have been studied, such as the one recently proposed by Safranek et al. \cite{safranek2019quantum}, which is a generalization of Boltzmann entropy to quantum physics, called ``observational'' entropy.   Gibbs's approach in classical physics starts with a density function in phase space, $\rho_{r,p} (\mathbf{r},\mathbf{p},t)$,  satisfying the  Liouville's equation, leading to the conclusion that the entropy associated with $\rho_{r,p} (\mathbf{r},\mathbf{p},t)$ is constant over time \cite{gibbs2014elementary}. From such density probability in phase space, we can obtain the marginal densities
\begin{align}
 \rho(\mathbf{r},t) & =|\psi(\mathbf {r},t)|^2=\int \diff^3\mathbf{p} \, \rho_{r,p}(\mathbf{r},\mathbf{p},t)
  \\
  \intertext{and}
 \rho_\mathrm{p}(\mathbf{p},t)& =|\phi(\mathbf {p},t)|^2=\int \diff^3\mathbf{r} \, \rho_{r,p}(\mathbf{r},\mathbf{p},t)\,,
\end{align}
which are the probability densities of the position and the momentum, respectively, where $\psi(\mathbf {r},t)$ is the probability amplitude and $\phi(\mathbf {p},t)$ is its Fourier transform. In quantum physics, the discretizaton and the finite nature of the volume $\diff V\equiv \diff^3 \mathbf{r} \diff^3 \mathbf{p} \ge \hbar^3$ is given by the uncertainty principle in position and momentum, as noted for example in \cite{jaynes1965gibbs}.
 However, despite much work, including \cite{wigner1932quantum, glauber1963coherent,husimi1940some}, it is unclear how to construct a relativistic density function in phase space.

 The entropic uncertainty inequality for the position and the momentum variables is
\begin{align}
 \mathrm{S}_t(\psi(\mathbf {r},t)) + \mathrm{S}_t(\phi(\mathbf {p},t)) \ge 3(1+\ln \piu),
 \label{eq:entropic-uncertainty-principle}
\end{align}
with the entropy given by
\begin{align}
\mathrm{S}_t(\psi(\mathbf {r},t)) = -\int \rho (\mathbf{r},t) \ln \rho (\mathbf{r},t) \,
 \diff^3\mathbf{r}\, ,
\label{eq:Relative-entropy}
\end{align}
and referred to as \emph{relative entropy}.  The entropic uncertainty was suggested by
\cite{hirschman1957note} and proved by \cite{babenko1961inequality} and \cite{beckner1975inequalities}. A good presentation is given by \cite{bialynicki1975uncertainty}, with a discrete version given in \cite{dembo1991information} and by Charles Peskin (unpublished communication, 2019). The entropic uncertainty is a tighter version of the inequality than the Heisenberg uncertainty principle based on variances, since $S(\psi )\leq \log {\sqrt {2\piu \eu V(|\psi |^2)}}$, where $V(|\psi |^2)$ is the variance of the probability density $\rho = |\psi |^2$. The Gaussian distribution obtains equality (minimum uncertainty) for both the entropic uncertainty and Heisenberg uncertainty principle. For a Gaussian distribution, the entropy increases with the variance, and if the variance in the position increases, the variance in the momentum decreases. We speculate that  as the entropy associated with the probability density $\rho(\mathbf{r},t)$ increases, the entropy associated with the probability density $\rho_p(\mathbf{p},t)$ decreases, and vice-versa. After all, it is true for the Gaussian distribution and more generally, the more information we have about the position of a particle, the less we know about the momentum, and vice-versa.

We then wonder whether in quantum physics a free particle probability density $\rho(\mathbf{r},t)$ disperses during the evolution, increasing the entropy. Is quantum physics equipped with a mechanism to increase entropy?   Is there a need of a law in quantum physics analogous to the second law of thermodynamics? If so, how would it impact the description of physical phenomena?

In this paper, we do adopt  the relative entropy formula given by~\eqref{eq:Relative-entropy}. However, we believe that other  possible  coherent measure of the information of a position over time can  be usefully studied.

\subsection{Our Contributions}
\label{sec:our-contributions}
As we have just discussed, we focus on the entropy associated with a given probability density. Such an entropy is a measure of the position information of the density function, where a uniform distribution, generated from plane waves probability amplitudes, has the smallest position information and the maximum entropy and a Dirac delta distribution has the largest position information and the minimum entropy. The probability amplitudes associated with these two distributions are not defined in Hilbert space but are idealizations of the two opposite limits of position information.

An entropy measure and a time interval naturally induce a partition of the Hilbert space into four sets. One contains the states for which the entropy is (i) increasing but not a constant and  another contains the states for which entropy is (ii) decreasing but not a constant. We establish an involution from set (i) onto set (ii) and show that if a state belongs to set (i) and evolves for a time interval, then the result of applying the conjugate operator to such evolved state will be a state in set (ii). However, this process of applying the conjugate operator to a state and evolving it according to the same Hamiltonian,  produces an anti-particle state, associated with the negative energy solution of a Hamiltonian.
Particles cannot be transformed into anti-particles due to conservation laws (such as charge conservation), unless the anti-particles are the particles themselves. Majorana fermions have such a property, and it is speculated that neutrinos are Majorana particles \cite{albert2014search}. In the boson ``family'' photons are their own anti-particles. Photons travel only at the speed of light and so their position  entropy is constant over time. Thus, photons cannot be in sets (i) or (ii).
Thus,  with the exception of Majorana particles, states in (i) cannot be transformed into states in (ii) and vice-versa.

Quantum  physical laws for fermions are described by Dirac's theory of free particles with a simplified approximation given by  \Schroedinger's Hamiltonian. The dispersion property of these Hamiltonians impacts the time evolution of the entropy. We show that  for a class of wave packets, the position entropy increases over time. The loss of position information occurs due to the dispersion properties of those Hamiltonians.

Thus, quantum physics is equipped with features that increases over time  the position entropy.  We mean entropy ``increases,''  as in the standard definition  that allows for the entropy to stay constant for any period of time.  The  physical reason why the entropy  associated with position information may increase  for fermions  is the dispersion property of the free-particle Hamiltonian,  coupled with conservation laws associated with the conjugate operator so that charged particles cannot become anti-particles. There are, however, scenarios where violations can occur. One such scenario, alluded to above, are  states characterizing Majorana particles where the conjugate operator  transforms a particle in itself, and the entropy can then decrease over a period of time.   Another scenario  are some superposition of stationary states of a Hamiltonian.  Stationary states are eigenstates of a Hamiltonian and thus,  their entropy remains constant over time.  A superposition of two stationary states will cause the entropy to oscillate over time. For example, that would be the case for any electron in an atom with the existence of two or more stationary states. It is not clear that such an example can occur, i.e., according to quantum field theory, the vacuum is a source of photon creation causing electrons at any other state to always transition to the ground state.
Other scenarios where the entropy decreases could occur in systems of multiple particles. We study simple scenarios of  collisions of a two-particle system, two fermions as well as two bosons. The entropy of each particle's wave packet  evolution alone increases due to dispersion.  For the pair of particles,  however,  as the distance between them is reduced during collision, and interference occurs,  an effect of reducing entropy kicks in. These two  evolution effects, each individual particle dispersing and the interference increasing,   compete to influence the entropy behavior during the evolution. We show that for slow speed collision, there is time for the dispersion effect of the Hamiltonian to dominate and the resulting entropy may increase over time. We also show that for faster-speed collision with shorter time for dispersion, the interference effect dominates, resulting in the entropy decreasing at some close distances.

We then postulate a law, \emph{the entropy law}, that the position information measured by the entropy increases over time.

The proposed entropy law restricts the set of physically-valid states. For example, the entire set (ii) is disallowed.  According to the law, a decay from the superposition of states to either a unique stationary state or to other forms of states would be needed. We speculate that free neutrinos, in a superposition of states, oscillate during flight, and may transform themselves according to this law. We speculate that high-speed collision of particles, such as $e^+ + e^- \rightarrow 2 \gamma$, produce new particles so that the entropy increases, while respecting conservation laws.

The paper is organized as follows. Section \ref{sec:quantum-review} is a review of selected topics in quantum physics focusing on technical results that relate directly to our postulate and its consequences. Section~\ref{sec:entropy-time-arrow}  reviews in detail the entropy description in quantum physics and  study the partition of the Hilbert space according to the entropy evolution.
Section~\ref{sec:entropy-law} develops the postulate that a time arrow is included in the laws of quantum theory and also examines the consequence for the case of a two-particle collision. Section~\ref{sec:conclusion} concludes the paper.

\section{Time Evolution and Coherent States }
\label{sec:quantum-review}
This section review some concepts already developed in quantum physics that we will directly use to formulate the role  of entropy for a one-particle quantum system.

\subsection{Time Evolution}

In order to fix the setting and the notation, we start by reviewing how time evolution is modeled in quantum physics. Quantum physics describes the evolution of a one-particle system as a state $\ket{\Psi_t}$ in Hilbert space that evolves over time parameterized by $t$. The evolution is governed by the motion equation and its solution
\begin{align}
 \iu \hbar {\frac {\partial }{\partial t}}\ket{\Psi_t}=H \ket{\Psi_t}, \thus \ket{\Psi_t}=e^{-\iu \frac{1}{\hbar} \int_0^t H(t') \diff t' } \ket{\Psi_0} ,
\label{eq:quantum-dynamics-II}
\end{align}
where $\hbar$ is the reduced Planck constant and $H$ is the Hamiltonian. The Hamiltonian is a Hermitian operator characterizing the evolution of the quantum state and it commutes with itself at different times: $[H(t),H(t')]=0$.
Since $H$ is Hermitian, the conjugate transpose of \eqref{eq:quantum-dynamics-II} is
\begin{align}
-\iu \hbar {\frac {\partial }{\partial t}}\bra{\Psi_t}=\bra{\Psi_t}H, \thus \bra{\Psi_t}=\bra{\Psi_0}e^{\iu\frac{1}{\hbar} \int_0^t H(t') \diff t' }\,.
\label{eq:quantum-dynamics-conjugate}
\end{align}

To avoid clumsy notation, we use the notation $\hat X$ for operators only when it is needed to disambiguate the operator $\hat X$ from the eigenvalue $X$.

One can represent the state $\ket{\Psi_t}$ in different bases according to the eigenstates of the operator of choice. For the position operator, $\hat{\mathbf{r}}$, the state description is $\Psi(\mathbf{r},t) $, or equivalently $ \bra{\mathbf{r}}\ket{\Psi_t}$, where $\ket{\mathbf{r}}$ are the eigenstates of the position operator, or $\hat{\mathbf{r}}\ket{\mathbf{r}}=\mathbf{r}\ket{\mathbf{r}}$.
\paragraph{Wave Function:} We will refer to $\Psi(\mathbf{r},t)$ as the wave function or the  probability amplitude. As we will elaborate, in the case of the \Schroedinger model it is a complex-valued function, while in the case of the Dirac model it is a complex-valued bi-spinor.
\paragraph{\Schroedinger Model:}
In the \Schroedinger equation for a non-relativistic particle the Hamiltonian is the sum of the kinetic and the potential energies, or $H={\mathbf {\hat p}^2}/{2m}+V(\mathbf {\hat r})$, and
\begin{align}
 \iu \hbar {\frac {\partial }{\partial t}}\Psi (\mathbf {r} ,t)=\iu \hbar {\frac {\partial }{\partial t}} \bra{\mathbf{r}}\ket{\Psi_t}=
 \mel{\mathbf{r}}{H}{\Psi_t}
=\left[{\frac {-\hbar ^{2}}{2m }}\nabla ^{2}+V(\mathbf {r})\right] \Psi (\mathbf {r} ,t),
\label{eq:Schroedinger}
\end{align}
where $m$ is the particle's mass, $\nabla$ is the gradient operator, $\nabla ^{2}=\nabla\cdot \nabla$ is the Laplacian operator, and  $\mathbf {\hat p}=-\iu \hbar \nabla$ in the space representation.
\paragraph{Dirac Model:}
 The relativistic equation for a fermion is
 \begin{align}
 (\iu \hbar \gamma ^{\mu }\partial _{\mu } -mc)\Psi =0,
 \label{eq:Dirac}
 \end{align}
 where $m$ is the mass of the particle at rest, $c$ is the speed of light, and $\Psi $ is a bi-spinor, a four-entry vector structure transforming as a bi-spinor, also referred to as a the Dirac spinor. The index $\mu=0,1,2,3$ varies over time and the three spatial coordinates with the special-relativity Minkowsky metric $\ket{-,+,+,+}$, and $\gamma ^{\mu }$ are $4 \times 4$ matrices satisfying the Clifford algebra. The Hamiltonian associated with \eqref{eq:Dirac} is
\begin{align}
 H =\left [ m c \gamma^0
+\iu \hbar\gamma^0 \vec{\gamma} \cdot \nabla \right ] ,
\label{eq:Hamiltonian-Dirac}
\end{align}
where $\vec{\gamma}=(\gamma^1, \gamma^2, \gamma^3)$.

 \paragraph{Born Rule:} The Born rule states that
 \begin{align}
 \rho(\mathbf{r} ,t)= \bra{\Psi_t}\ket{\mathbf{r}}\bra{\mathbf{r}}\ket{\Psi_t}= \abs{\Psi(\mathbf{r},t)}^2
 \label{eq:probability-density}
 \end{align}
 is the probability density function for both complex-valued \Schroedinger waves and for Dirac spinors, and therefore $1=\int \! \rho(\mathbf{r} ,t) \diff^3 \mathbf{r} $. For the Dirac spinors $|\Psi(\mathbf{r},t)|^2=\Psi^{\dag}(\mathbf{r},t) \Psi(\mathbf{r},t)$, where $\Psi^{\dag}(\mathbf{r},t) $ is the Hermitian of $\Psi$.

\subsection{Fourier Space: Phase Velocity, Group Velocity, and the Hessian}

A standard Fourier method transforms a function from the spatial basis representation $\ket{\mathbf{r}}$ to the momentum basis representation $\ket{\mathbf{p}}$ (or $\ket{\mathbf{k}}$, since $\mathbf {p}=\hbar \mathbf {k}$) and vice-versa. In particular, $\bra{\mathbf{r}}\ket{\mathbf{k}}=\eu^{\iu \mathbf{k}\cdot \mathbf{r}}$. Extending the standard spatial Fourier method to time, we obtain the Fourier transform from a space-time representation to an energy-momentum representation. We adopt the special relativity metric $\ket{-,+,+,+}$ for the scalar product of vectors $(t,\mathbf{r})$ and $(\omega,\mathbf{k})$. More precisely, we write the inverse Fourier transform with Minkowsky metric, which is a four dimensional transformation, as
\begin{align}
\Psi (\mathbf {r} ,t)&={\frac {1}{({\sqrt {2\piu }})^{4}}}\int \! \diff \omega \int \Phi (\omega, \mathbf {k} )\, \eu^{\iu(-\omega t+\mathbf {k} \cdot \mathbf {r} )}\diff^{3} \mathbf {k} \, ,
\label{eq:Fourier-Method-Minkowsky}
\end{align}
where $\Phi (\omega, \mathbf {k} )=\bra{\omega, \mathbf {k}}\ket{\Psi_t}$. Associating the energy with $E=\hbar \omega$, the Fourier space is the energy-momentum space, that is, $(E,\mathbf {p})=\hbar (\omega,\mathbf {k})$. Note that the energy values are eigenvalues of the Hamiltonian and so the integral \eqref{eq:Fourier-Method-Minkowsky} in the variable $\omega$ can have regions of finite sums.

The free particle Hamiltonians are $H_{\mathcal{S}} = \left[{\frac {-\hbar ^{2}}{2m }}\nabla ^{2}\right]$ and $H_{\mathcal{D}} =\left [ m c \gamma^0 +\iu \hbar\gamma^0 \vec{\gamma} \cdot \nabla \right ]$ for the \Schroedinger equation \eqref{eq:Schroedinger} and the Dirac equation \eqref{eq:Dirac}, respectively. These are descriptions in position-time space, and we can also write them in Fourier space. Both Hamiltonians are functions of the momentum operator and therefore can be diagonalized in the $\ket{\mathbf {k}}$ basis (spatial Fourier domain),  as derived in Appendix~\ref{eq:Schroedinger-momentum} and Appendix~\ref{sec:Dirac-momentum}, to obtain respectively
\begin{align}
\omega^{\cal S}(\mathbf{k}) & =\frac{\hbar}{2m} \matrixsym{k}^2&& \text{(\Schroedinger equation) \quad and}
\\
\omega^{\cal D}(\mathbf{k})& =\pm c \sqrt{ \matrixsym{k}^2+\frac{m^2}{\hbar^2} c^2} &&
\text{(Dirac equation).}
\label{eq:Fourier-Hamiltonians}
\end{align}

 The group velocity becomes respectively
 \begin{align}
\mathbf{v_g}^{\cal S}(\mathbf{k})=\nabla_{\mathbf {k}} \omega^{\cal S}(\mathbf{k}) & =\frac{\hbar}{m} \mathbf {k} && \text{(\Schroedinger equation) \quad and}
\\
\mathbf{v_g}^{\cal D}(\mathbf{k})=\nabla_{\mathbf {k}}\omega^{\cal D}(\mathbf{k})& =\pm c \frac{\mathbf {k}}{\sqrt{ \matrixsym{k}^2+\frac{m^2}{\hbar^2} c^2}}=
\pm \frac{\hbar}{m}\frac{\mathbf {k}}{\sqrt{(\frac{\hbar \matrixsym{k}}{m c})^2+1}} &&
\text{(Dirac equation).}
\label{eq:Fourier-group-velocity}
 \end{align}

In Section~\ref{sec:entropy-time-arrow}, we will use the Taylor expansions of equations \eqref{eq:Fourier-Hamiltonians} up to the second order, see \eqref{eq:w-dispersion}, and so we will need the Hessians, respectively
\begin{align}
{\cal H}_{ij}^{\cal S}(\mathbf{k}) &=\frac{\partial^2 \omega^{\cal S}(\mathbf{k}) }{\partial \matrixsym{k}_i \partial \matrixsym{k}_j} =\frac{\hbar}{m} \deltau_{i,j} && \text{(\Schroedinger equation) \quad and}
\\
{\cal H}_{ij}^{\cal D}(\mathbf{k})&=\frac{\partial^2 \omega^{\cal D}(\mathbf{k})}{\partial \matrixsym{k}_i \partial \matrixsym{k}_j}=\pm c \left [ \frac{\deltau_{i,j}}{\left ( \matrixsym{k}^2+\frac{m^2}{\hbar^2} c^2\right)^{\frac{1}{2}}}-\frac{\matrixsym{k}_i \matrixsym{k}_j}{\left ( \matrixsym{k}^2+\frac{m^2}{\hbar^2} c^2\right)^{\frac{3}{2}}}\right ] &&
\\
&= \pm \frac{\hbar}{m} \left (1+\left (\frac{\hbar \matrixsym{k}}{m c}\right)^2\right)^{-\frac{3}{2}}\left [ \deltau_{i,j}\left (1+\left (\frac{\hbar \matrixsym{k}}{m c}\right)^2 \right) -\left (\frac{\hbar \matrixsym{k}_i}{m c}\right) \left (\frac{\hbar \matrixsym{k}_j}{m c}\right)\right ] &&  \text{(Dirac equation)}.
\label{eq:Fourier-group-Hessian}
\end{align}

A Hessian gives a measure of dispersion of the wave. The Hessian for the \Schroedinger equation is the identity matrix scaled by ${\hbar}/{m}$. This matrix is positive and the larger is the mass of the particle, the smaller are the eigenvalues and the dispersion. For the Dirac equation the eigenvalues are
\begin{align}
\lambda_1^D&=\pm \frac{\hbar}{m}\left (1+\left (\frac{\hbar \matrixsym{k}}{m c}\right)^2\right)^{-\frac{3}{2}}=\pm \hbar \frac{m^2}{\left (m^{2}+\mu^2(\matrixsym{k})\right)^{\frac{3}{2}}}\, \text{ \qquad and}
\\
\lambda_{2,3}^D& =\pm \frac{\hbar}{m} \left (1+\left (\frac{\hbar \matrixsym{k}}{m c}\right)^2\right)^{-\frac{1}{2}}=
\pm \hbar \frac{1}{(m^2+\mu^2(\matrixsym{k}))^{\frac{1}{2}} }\, ,
\end{align}
where $\mu(\matrixsym{k})={\hbar \matrixsym{k}}/{c}$ is a measure of the kinetic energy in mass units and the second eigenvalue has multiplicity two. Thus, for both equations the Hessian is positive definite for positive energy solutions. For $\lambda_{2,3}^\mathcal{D}$, the larger is the mass of the particle, the smaller are the eigenvalues and the dispersion. However, for $\lambda_1^\mathcal{D}$ and for the mass values  where  $m \le \sqrt{2} \mu(\matrixsym{k})$,
the eigenvalues and dispersion  increase as the mass increases.

\subsection{Antiparticles and Conjugate Solutions}

 The eigenvalues of the Dirac energy-momentum matrix equation, as derived in Appendix~\ref{sec:Dirac-momentum}, are $\omega(\matrixsym{k}^2)=\pm c \sqrt{ \matrixsym{k}^2+\frac{m^2}{\hbar^2} c^2}$, each in multiples of two.
There are four eigenvectors associated with each linear equation and they are typically described in terms of two pairs of 2D spinors
\begin{align}
 \chi^{\pm R}(\omega, \mathbf {k}) =\sqrt{\omega+c(\matrixsym{\sigma}\cdot \matrixsym{k})} \, \xi^{\pm}
 \quad {\rm and}\quad
 \chi^{\pm L}(\omega, \mathbf {k}) =\sqrt{\omega-c(\matrixsym{\sigma}\cdot \matrixsym{k})} \, \xi^{\pm}\, ,
 \label{eq:chiRL}
\end{align}
where $\xi^{\pm}$ are two normalized vectors in 2D, with $\xi^+=\begin{pmatrix} 1\\0 \end{pmatrix} $ for spin up and $\xi^-=\begin{pmatrix} 0\\1 \end{pmatrix} $ for spin down. In this representation, the four orthogonal eigenvector solutions $\mu^{\pm} (\omega , \mathbf {k}) $ and $ \nu^{\pm} (\omega , \mathbf {k}) $ of the Dirac matrix equation in energy-momentum space are
 \begin{align}
  \omega(\matrixsym{k}^2)=\sqrt{c^2 \matrixsym{k}^2+\frac{m^2}{\hbar^2} c^4}, \thus
 \mu^{\pm} (\omega , \mathbf {k}) &=\begin{pmatrix}
 \sqrt{\omega-c(\matrixsym{\sigma}\cdot \matrixsym{k})} \, \xi^{\pm}
 \\[0.9ex]
 \sqrt{\omega+c(\matrixsym{\sigma}\cdot \matrixsym{k})} \, \xi^{\pm}
 \end{pmatrix}
 =\begin{pmatrix}
 \chi^{\pm L}(\omega, \mathbf {k})
 \\
 \chi^{\pm R}(\omega, \mathbf {k})
 \end{pmatrix}\, ,\quad \text{and}&
 \\[1.0ex]
 \omega(\matrixsym{k}^2)=- \sqrt{c^2 \matrixsym{k}^2+\frac{m^2}{\hbar^2} c^4}, \thus
 \nu^{\pm} (\omega , \mathbf {k}) &=\iu \begin{pmatrix}
\sqrt{\omega+c(\matrixsym{\sigma}\cdot \matrixsym{k})} \, \xi^{\pm}
 \\[0.9ex]
 -\sqrt{\omega-c(\matrixsym{\sigma}\cdot \matrixsym{k})} \, \xi^{\pm}
 \end{pmatrix}
 =\iu \begin{pmatrix}
\chi^{\pm R}(\omega, \mathbf {k})
 \\
 - \chi^{\pm L}(\omega, \mathbf {k})
 \end{pmatrix} \, .&
 \label{eq:Dirac-Solutions-Fourier}
\end{align}
 The global phase $\iu=e^{\iu \frac{\piu}{2}}$ for the negative energy eigenvectors is arbitrary and is introduced only for the convenience of the manipulations that follow. Thus, the four orthogonal solutions \eqref{eq:Dirac-Solutions-Fourier} in space-time are
 \begin{align}
 \Psi^{\pm}_t(\mathbf {r} ,t)
&
= \frac {1}{({\sqrt {2\piu }})^{3}} \int \mu^{\pm} \left (\sqrt{c^2 \matrixsym{k}^2+\frac{m^2}{\hbar^2} c^4}, \mathbf {k} \right)
\eu^{-\iu t\, \sqrt{c^2 \matrixsym{k}^2+\frac{m^2}{\hbar^2} c^4}} \eu^{\iu \mathbf {k}\cdot \mathbf {r}}\diff^3\mathbf {k} \, , \qquad \text{and}
\\
\Psi^{\pm}_{-t}(\mathbf {r} ,t)
&= \frac {1}{({\sqrt {2\piu }})^{3}} \int \nu^{\pm} \left (-\sqrt{c^2 \matrixsym{k}^2+\frac{m^2}{\hbar^2} c^4},\mathbf {k} \right )
\eu^{\iu t\, \sqrt{c^2 \matrixsym{k}^2+\frac{m^2}{\hbar^2} c^4}} \eu^{\iu \mathbf {k}\cdot \mathbf {r}} \diff^3\mathbf {k}\, ,
\label{eq:Dirac-Fourier-timespace-constrained}
 \end{align}
 where we assign the indices $t$ and $-t$ to each pair of solutions to indicate the sign of $t$ in the phase of the exponential term. The $-t$ solutions will yield the two antiparticle solutions as follows. First, consider the adjoint of $\Psi^{\pm}_{-t}(\mathbf {r} ,t) $
 \begin{align}
 \overline \Psi_{-t}^{\pm}(\mathbf {r} ,t) &=(\Psi^{\pm}_{-t})^{\dag}(\mathbf {r} ,t) \gamma^0
  = \frac {1}{({\sqrt {2\piu }})^{3}} \int \diff^3\mathbf {k}\, {(\nu^{\pm})}^{\dag} (\omega(\matrixsym{k}^2) , \mathbf {k})\, \gamma^0 \,
\eu^{-\iu t \sqrt{c^2 \matrixsym{k}^2+\frac{m^2}{\hbar^2} c^4}} \eu^{-\iu \mathbf {k}\cdot \mathbf {r}}\, ,
 \\
 \intertext{and after using the adjoint representation $\overline{\nu}^{\pm}(\mathbf {k}) ={(\nu^{\pm})}^{\dag} (\omega(\matrixsym{k}^2) , \mathbf {k}) \gamma^0$ , }
 \overline \Psi_{-t}^{\pm}(\mathbf {r} ,t) &=\frac {1}{({\sqrt {2\piu }})^{3}} \int \overline{\nu}^{\pm}(\mathbf {k})
\,
\eu^{-\iu t \sqrt{c^2 \matrixsym{k}^2+\frac{m^2}{\hbar^2} c^4}} \eu^{-\iu \mathbf {k}\cdot \mathbf {r}}\diff^3\mathbf {k}\, ,
\\
\intertext{and after changing variables (reversing the momentum): $\mathbf {k}'=- \mathbf {k}$\,,}
 \overline \Psi_{-t}^{\pm}(\mathbf {r} ,t) &=-\frac {1}{({\sqrt {2\piu }})^{3}} \int \overline{\nu}^{\pm} (-\mathbf {k}')
\eu^{-\iu t \sqrt{c^2 \matrixsym{k'}^2+\frac{m^2}{\hbar^2} c^4}} \eu^{\iu \mathbf {k}'\cdot \mathbf {r}} \diff^3\mathbf {k}'\, .
 \end{align}
The antiparticle solution solution is then a bi-spinor solution
 \begin{align}
\Psi_{A}^{\pm}(\mathbf {r} ,t) &\equiv (\overline \Psi_{-t}^{\pm}(\mathbf {r} ,t))^{\tran}=\frac {1}{({\sqrt {2\piu }})^{3}} \int \nu_A^{\pm}( \mathbf {k})
\eu^{-\iu t \sqrt{c^2 \matrixsym{k}^2+\frac{m^2}{\hbar^2} c^4}} \eu^{\iu \mathbf {k}\cdot \mathbf {r}}\diff^3\mathbf {k},
\label{eq:adjoint-solution-anti-particle}
 \end{align}
 where in the Weyl chiral basis
 \begin{align}
 (\nu_A^{\pm}(\mathbf {k}))^{\tran} & =-\overline{\nu}^{\pm} (-\mathbf {k})=\iu \begin{pmatrix}
\sqrt{-c \left (\sqrt{\matrixsym{k}^2+\frac{m^2}{\hbar^2} c^2}+(\matrixsym{\sigma}\cdot \matrixsym{k})\right)} \xi^{\pm}
\\[2.8ex]
 -\sqrt{-c\left (\sqrt{\matrixsym{k}^2+\frac{m^2}{\hbar^2} c^2}-(\matrixsym{\sigma}\cdot \matrixsym{k})\right)} \xi^{\pm}
 \end{pmatrix}^{\dag} \gamma^0
 =\begin{pmatrix}
 \chi^{\pm L}(\mathbf {k})
\\
 -\chi^{\pm R}(\mathbf {k})
 \end{pmatrix}^\tran
 \label{eq:anti-particle-representation}
 \end{align}
 with $\chi^{\pm R}(\mathbf {k}) =\sqrt{c \left (\sqrt{\matrixsym{k}^2+\frac{m^2}{\hbar^2} c^2}+(\matrixsym{\sigma}\cdot \matrixsym{k})\right)} \xi^{\pm} $ and $\chi^{\pm L}(\mathbf {k}) =\sqrt{c \left (\sqrt{\matrixsym{k}^2+\frac{m^2}{\hbar^2} c^2}-(\matrixsym{\sigma}\cdot \matrixsym{k})\right)} \xi^{\pm} $.
 The Dirac spinors $\Psi_{A}^{\pm}(\mathbf {r} ,t)$ have the same probability density functions as $\Psi^{\pm}_{-t}(\mathbf {r} ,t)$, but with the charge and the momentum reversed. They are referred to as antiparticle waves or fields, when they are promoted to operators.
 Thus, the time evolutions of the momentum solutions are given by
 \begin{align}
 \mu^{\pm} (\mathbf {k},t)& = \mu^{\pm} (\mathbf {k})\,
\eu^{-\iu \sqrt{c^2 \matrixsym{k}^2+\frac{m^2}{\hbar^2} c^4}\, t}
=\begin{pmatrix}
 \chi^{\pm L}(\mathbf {k})
 \\
 \chi^{\pm R}(\mathbf {k})
 \end{pmatrix}\, \eu^{-\iu \, t\, \sqrt{c^2 \matrixsym{k}^2+\frac{m^2}{\hbar^2} c^4}}\,  \qquad \text{and}
\\
\nu_A^{\pm}( \mathbf {k},t) &= \nu_A^{\pm}( \mathbf {k})\,
\eu^{-\iu \sqrt{c^2 \matrixsym{k}^2+\frac{m^2}{\hbar^2} c^4}\, t}
= \begin{pmatrix}
 \chi^{\pm L}(\mathbf {k})
 \\
 - \chi^{\pm R}(\mathbf {k})
 \end{pmatrix} \, \eu^{-\iu \, t \, \sqrt{c^2 \matrixsym{k}^2+\frac{m^2}{\hbar^2} c^4}} \, .
 \end{align}

 \subsubsection{A Remark on the Conjugation Operation}
\begin{figure}
 \centering
 \includegraphics[width=0.95\linewidth]{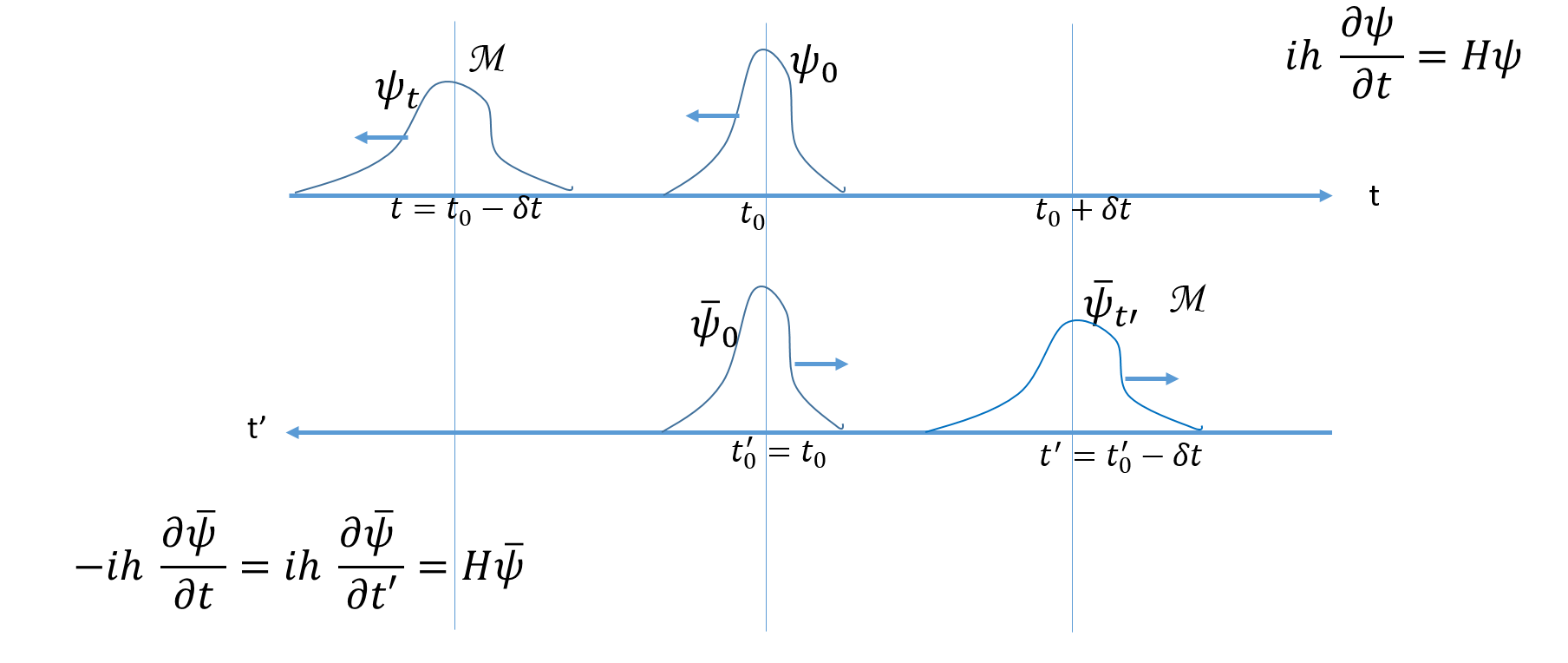}
 \caption{A quantum representation equivalence. (i) Top figure axis $t$: A one-dimensional wave packet moving backward in time. It is described by $\Psi^{\pm}_{-t}(\mathbf {r} ,t)$ in \eqref{eq:Dirac-Fourier-timespace-constrained} with negative energy, $\omega<0$, and the time evolution term $\eu^{-\iu \omega t}=\eu^{\iu |\omega| t}$. This is equivalent to $\eu^{-\iu |\omega| (-t)}$, i.e., a positive energy solution, $|\omega|$, moving backward in time ($-t$).
 (ii) Bottom figure axis $t'$: The motion equation for the adjoint differs only in the sign of $t$, and by defining $t'=t_0-t$ we represent each axis opposite to each other, i.e., when $t'$ decreases, $t$ increases. The adjoint of the solution in (i) is $\overline \Psi_{-t}^{\pm}(\mathbf {r} ,t) =(\Psi^{\pm}_{-t})^{\dag}(\mathbf {r} ,t) \gamma^0 $ (and $\Psi_{A}^{\pm}(\mathbf {r} ,t) \equiv (\overline \Psi_{-t}^{\pm}(\mathbf {r} ,t))^{\tran}$ as in \eqref{eq:adjoint-solution-anti-particle}), i.e., described by the time evolution $\eu^{\iu \omega t}=\eu^{-\iu |\omega| t}$, a positive energy solution moving forward in time. Both solutions, $\Psi^{\pm}_{-t}(\mathbf {r} ,t)$ and $\Psi_{A}^{\pm}(\mathbf {r} ,t)$, have the same magnitude squares and therefore they yield the same probability densities. The equivalence can be generally stated as follows:
 \emph{There is an equivalence between describing a particle by the probability amplitude and its motion equation, and describing a particle by its conjugate and the evolution by the adjoint of the motion equation.} A similar statement is captured by the Feynman-Stueckelberg interpretation.
 }
 \label{fig:anti-particles}
 \end{figure}

 The conjugation operation, also called charge conjugation, on a state $\ket{\psi}$ is  defined  by ${\cal C} \ket{\psi}=\ket{\overline{\psi}} $, where ${\cal C}$ is the Hermitian and unitary conjugate operator. It is known (e.g., \cite{1968DiracLectures}) that the sign of all the quantum charges (electric, lepton flavor, quark  flavor)  of $\ket{\overline{\psi_t}}$ are reversed to those of $\ket{\psi_t}$. Thus, antiparticles will have the charge reversed with respect to the state that is conjugated to yield the antiparticle.  In the above calculations, the state $\Psi^{\pm}_{-t}(\mathbf {r} ,t)$ was conjugated to yield $\Psi^{\pm}_{A}(\mathbf {r} ,t)$.  The states $\Psi^{\pm}_{t}(\mathbf {r} ,t)$ with the same charge sign as $\Psi^{\pm}_{-t}(\mathbf {r} ,t)$   will have a charge opposite to $\Psi^{\pm}_{A}(\mathbf {r} ,t)$. The Feynman-Stueckelbert interpretation of antiparticles  considers $\Psi^{\pm}_{-t}(\mathbf {r} ,t)$  and $\Psi^{\pm}_{A}(\mathbf {r} ,t)$ to be equivalent representations of the particle. In Figure~\ref{fig:anti-particles} we illustrate this quantum-representation equivalence and elaborate on it further. In this equivalence,  the choice for a particle representation is the one with positive energy and with the time going forward.

\subsection{Coherent States and Ladder Operators}
 The ladder operators per spatial dimension $\hat a_i $, $ i=1,2,3$ and their adjoints $\hat a^{\dagger}_i$, $ i=1,2,3$ are defined as
\begin{align}
 \hat a_i=\frac{1}{\lambda_0}\mathbf{\hat {r}_i}+ \iu \frac{\lambda_0}{\hbar}\mathbf{\hat {p}_i}\quad \textrm{and} \quad
 \hat a^{\dagger}_i=\frac{1}{\lambda_0}\mathbf{\hat {r}_i}- \iu \frac{\lambda_0}{\hbar}\mathbf{\hat {p}_i},
\end{align}
where the spatial index is $i=1,2,3$, and there is a natural length parameter $\lambda_0=\sqrt{{2\hbar}/{ m \omega_0}}$ associated with a natural frequency $\omega_0$. Similarly to the commutation properties for the position and momentum operators $[\mathbf{\hat {r}}_i, \mathbf{\hat {p}}_j ]=\iu \frac{\hbar}{2} \deltau_{ij}$, we have $[\hat a_i, \hat a^{\dagger}_j]=\deltau_{ij}$.

 Coherent states fir spatial dimensions $\ket{\alpha_i} $, $ i=1,2,3 $ are the eigenstates of the ladder operator $\hat a_i $, $ i=1,2,3$ and they attain the minimum uncertainty principle. Coherent states were first derived by \Schroedinger \cite{schroedinger1930a} and  are represented as
\begin{align}
 \hat a_i \ket{\alpha_i} &= \alpha_i \ket{\alpha_i}\, ,
 \\
\intertext{and therefore, up to a global phase,}
\ket{\alpha_i}&=\eu^{|\alpha_i|}\sum_{n_i=0}^{\infty}\frac{\alpha_i^{n_i}}{\sqrt{n_i!}} \ket{n_i}\,  \qquad \text{and}
 \\
 \bra{\mathbf{r}_i}\ket{\alpha_i}&=
 \normalx{\mathbf {r}_i}{0}{\lambda_0^2}\, \eu^{ \iu k_i \mathbf{r}_i}\, ,
 \label{eq:coherent-states}
\end{align}
where $\alpha_i=\iu \lambda_0 k_i$, $n_{i}$ is a positive integer, and $\ket{n_i}$ represents the state with the specific number $n_i$. The number operators $\hat n_i=a_i^{\dagger} a_i$ satisfy $\hat n_i \ket{n_i}=n_i \ket{n_i} $. The amplitudes $ \bra{\mathbf{r}_i}\ket{\alpha_i}$ are spatially localized and are in Hilbert space. They also represent the eigenfunctions of the ground state of the quantum harmonic oscillator  with natural frequency $\omega_0$, which models the electromagnetic field Hamiltonian. Furthermore, they are also an overcomplete representation of all the functions in Hilbert space as we vary $\lambda_0$ and $k_i$. This set of functions differs from the Fourier basis $\eu^{\iu k_i r_i}$ as their real components decay over space. Such a property places them in Hilbert space and will be exploited in our development.

\subsubsection{Bosons and Fermions}

The ladder operators described above yield the number operator, $\matrixsym{\hat N} = \sum_{i=1}^{3} \hat n_i= \sum_{i=1}^{3}\hat a^{\dagger}_i \hat a_i $. States are then created from the lowest energy state, $\ket{0}$, using the creation operator, as
\begin{align}
 \ \ket{n_x, n_y, n_z}= \frac{1}{\sqrt{n_z!}}(\hat a^{\dagger}_z)^{n_z} \frac{1}{\sqrt{n_y!}}(\hat a^{\dagger}_y)^{n_y} \frac{1}{\sqrt{n_x!}}(\hat a^{\dagger}_x)^{n_x} \ket{0}\, ,
\end{align}
where $n_x=n_1$, $ n_y=n_2$, and $ n_z=n_3$.

The description above is valid for bosons. In the following, $i= 1, 2, 3$. When describing fermions, the ladder operators $ \{\hat b_i, \hat b^{\dagger}_i\}=1 $ satisfy instead the anti-commuting rule (or Poisson brackets) $\{\hat b_i, \hat b^{\dagger}_i\}=1$ including with themselves $\{\hat b_i, \hat b_i\}=\{\hat b_i^{\dagger}, \hat b^{\dagger}_i\}=0 $ . Thus, $\hat b_i\hat b_i=\hat b^{\dagger}_i\hat b^{\dagger}_i=0$. The number operator per dimension is also given by $\hat N_i=\hat b^{\dagger}_i \hat b_i$. However, while $\hat a_i\hat a^{\dagger}_i=\mathrm{I} + \hat N_i$ for bosons, we have $\hat b_i\hat b^{\dagger}_i=\mathrm{I}-\hat N_i$ for fermions.

Ladder operators and their eigenstates (coherent states) belong to the foundations of quantum field theory and will be used here as the foundation for a particle description.

\section{Entropy in Quantum Physics}
\label{sec:entropy-time-arrow}

As discussed in the introduction, the information  measure of  the  density function $\rho(\mathbf {r} ,t)=\Psi^{\dag} (\mathbf {r} ,t)\, \Psi (\mathbf {r} ,t) $ is given by the relative entropy \eqref{eq:Relative-entropy}.
Note that we are focusing on the entropy associated with the position information,  and referring to it by the general term \textit{entropy}.

Let us define the adjoint of the state $\ket{\psi_t}$ by $\bra{\overline{\psi_t}}=\bra{\psi_t}\gamma^0$, i.e., the conjugate, transpose, multiplied by $\gamma^0$.
We will use the following simple fact throughout the paper, and therefore state it as a lemma.
\begin{lemma}
\label{lemma:Entropy-conjugate}
The state $\ket{\psi_t}$, the conjugate $\ket{\psi_t^*}$,  and the adjoint $\bra{\overline{\psi_t}}=\bra{\psi_t}\gamma^0$, all have the same entropy.
\end{lemma}
\begin{proof}
 The probability densities for $\ket{\psi_{t}}$, $\ket{\psi_{t}^*}$, and $\bra{\overline{\psi_t}}=\bra{\psi_t}\gamma^0$
are $\bra{\psi_{t}}\ket{\psi_{t}}=\psi^{\dagger}(\mathbf {r},t)\, \psi(\mathbf {r},t)$, $\bra{\psi_{t}^*}\ket{\psi_{t}^*}=\left(\bra{\psi^*_{t}}\ket{\psi^*_{t}}\right)^*=\left(\psi^{\tran}(\mathbf {r},t)\, \psi^*(\mathbf {r},t) \right)^*=\psi^{\dagger}(\mathbf {r},t)\, \psi(\mathbf {r},t)$ and $\bra{\overline{\psi_t}}\ket{\overline{\psi_t}}=\psi^{\dagger}(\mathbf {r},t) \gamma^0 \gamma^0 \psi(\mathbf {r},t)  = \psi^{\dagger}(\mathbf {r},t)\psi(\mathbf {r},t) $, respectively. As these  probability densities are equal, so are their associated entropies.
\end{proof}

\subsection{Stationary States}
The stationary states are the eigenstates of the Hamiltonian.

\begin{lemma}
\label{lemma:stationary-entropy}
Let $H$ be a time-invariant Hamiltonian and $\ket{\psi_{t}}$ its eigenstate. Then the entropy during the evolution of $\ket{\psi_{t}}$ is time invariant.
\end{lemma}
\begin{proof}
A quantum eigenstate of the Hamiltonian with eigenvalue $E$ is described as a wave function $\Psi_{E}(\mathbf{r})=\bra{\mathbf{r}}\ket{\psi^E}$ and evolves as $\Psi(\mathbf{r},t)=\bra{\mathbf{r}}\ket{\psi^E_t}=\Psi_{E}(\mathbf{r}) \, \eu^{-\iu \frac{E}{\hbar} t}$. Thus the probability density is $\rho(\mathbf{r},t)=|\Psi_{E}(\mathbf{r})|^2$. It is time invariant and so is the associated entropy. .
\end{proof}
 Stationary states for the Dirac and \Schroedinger equation for free particles include all the plane wave solutions $\Psi(\mathbf{r},t)=A\, \eu^{\iu (\mathbf{k} \cdot \mathbf{r}-\omega t)} $, though they are not elements of the Hilbert space due to the lack of normalization.

 \subsection{Time-Evolution of Waves and a Dispersion Transform}

 We now consider initial solutions that are localized in space of the $\psi_{\mathbf {k}_0}(\mathbf {r}-\mathbf {r}_0)=\psi_0(\mathbf {r}-\mathbf {r}_0) \, \eu^{\iu \mathbf {k}_0\cdot \mathbf {r}}$, where $\mathbf{r}_0$ is the mean value of $\mathbf {r}$ according to the probability distribution $\rho(\mathbf {r})=|\psi_0(\mathbf {r})|^2$, and the phase term $\eu^{\iu \mathbf {k}_0\cdot \mathbf {r}}$ gives the momentum shift of $\mathbf {k}_0$. Assume that the variance of $\mathbf {r}$, $\sigma^2 =\int \diff^3\mathbf {r}\, (\mathbf {r}-\mathbf {r}_0)^2 \rho(\mathbf {r}) $, is finite.  $\psi_{\mathbf {k}_0}(\mathbf {r}-\mathbf {r}_0)$ evolves according to the given Hamiltonian with a dispersion relation $\omega(k)$. We can represent the initial state in the momentum space as $ \Phi_{\mathbf {r}_0}(\mathbf {k}-\mathbf {k}_0)= \Phi_{0}(\mathbf {k}-\mathbf {k}_0) \, \eu^{-\iu (\mathbf {k}-\mathbf {k}_0)\cdot \mathbf {r}_0}$, where $\Phi_{0}(\mathbf {k})$ is the Fourier transform of $\psi_0(\mathbf {r})$. By the Fourier properties, $\rho(\mathbf {k})=|\Phi_{\mathbf {r}_0}(\mathbf {k}-\mathbf {k}_0)|^2$ also has a finite variance, $\sigma_{\mathbf {k}}^2$,  with the mean in the center momentum $\mathbf {k}_0$. Evolving the wave function in the momentum space according to the dispersion relation and taking the inverse Fourier transform, we get
\begin{align}
 \psi(\mathbf {r}-\mathbf {r}_0,t) &={\frac {1}{({\sqrt {2\piu }})^{3}}}\int \Phi_{\mathbf {r}_0} (\mathbf {k}-\mathbf {k}_0) \eu^{-\iu \omega(\mathbf {k}) t} \eu^{\iu \mathbf {k} \cdot \mathbf {r} }\diff^{3}\mathbf {k}\,.
 \label{eq:time-evolution-psi-general}
\end{align}
Since $\Phi_{\mathbf {r}_0}(\mathbf {k}-\mathbf {k}_0)$ fades away exponentially from $\mathbf {k}=\mathbf {k}_0$, we expand the dispersion formula in Taylor series
\begin{align}
 \omega(\mathbf {k}) & \approx
 \omega(\mathbf {k}_0)+\nabla_{\mathbf {k}}\omega(\mathbf {k})\Big |_{\mathbf {k}_0}\! \cdot (\mathbf {k}-\mathbf {k}_0)+\frac{1}{2}
 \frac{\partial^2 \, \omega(\mathbf{k} ) }{\partial \matrixsym{k}_i \partial \matrixsym{k}_j} \Big |_{\mathbf {k}_0}
 (\matrixsym{k}_i-(\matrixsym{k}_0)_i)(\matrixsym{k}_j-(\matrixsym{k}_0)_j) +\hdots
 \\
 &\approx
 \mathbf{\mathbf{ v_p}}(\mathbf {k}_0) \cdot \mathbf {k}_0+\mathbf{v_g}(\mathbf {k}_0) \cdot (\mathbf {k}-\mathbf{k}_0)+
 \frac{1}{2}(\mathbf {k}-\mathbf {k}_0)^{\tran} {\cal H}(\mathbf {k}_0) \,
 (\mathbf {k}-\mathbf {k}_0) +\hdots\,,
 \label{eq:w-dispersion}
\end{align}
where $\mathbf{\mathbf{ v_p}}(\mathbf {k}_0)$, $\mathbf{ \mathbf{v_g}}(\mathbf {k}_0)$, and $ {\cal H}(\mathbf {k}_0) $ are the phase velocity, the group velocity, and the Hessian of the dispersion relation $\omega(\mathbf {k})$, respectively. Then, inserting this approximation back into \eqref{eq:time-evolution-psi-general}, we get
\begin{flalign}
 \psi_{\mathbf {k}_0}(\mathbf {r}-\mathbf {r}_0-\mathbf{v_g}(\mathbf {k}_0) t,t) & \approx
 \frac{\eu^{\iu \mathbf {k}_0\cdot (\mathbf {r}-\mathbf{\mathbf{ v_p}} \, t)}}{Z}\, {\frac {1}{({\sqrt {2\piu }})^{3}}} &
 \\
 &\quad \times \int \Phi_{0}(\mathbf {k}-\mathbf {k}_0)\, \eu^{-\iu \frac{t}{2}(\mathbf {k}-\mathbf {k}_0)^{\tran}\, {\cal H}(\mathbf {k}_0) \, (\mathbf {k}-\mathbf {k}_0) } \, \eu^{\iu (\mathbf {k}-\mathbf {k}_0) \cdot (\mathbf {r}- \mathbf {r}_0 -\mathbf{v_g}(\mathbf {k}_0) t) }\diff^{3}\mathbf {k}&
 \\
 & =
 \frac{1}{Z} \psi_{\mathbf {k}_0}(\mathbf {r}- (\mathbf {r}_0 +\mathbf{v_g}(\mathbf {k}_0) t)) \ast \normalx{\mathbf {r}}{\mathbf{r}_0+\mathbf{v_g}(\mathbf {k}_0) t}{\iu t {\cal H}(\mathbf {k}_0) },&
 \label{eq:time-evolution-psi-dispersion}
\end{flalign}
where $\ast$ denotes a convolution, $Z$ normalizes the amplitude, and $\mathcal{N}$ denotes a normal distribution. We can interpret this evolution as describing a wave moving with phase velocity $\mathbf{\mathbf{ v_p}}(\mathbf {k}_0)$, group velocity $\mathbf{v_g}(\mathbf {k}_0)$, and being blurred by a time varying  complex only valued symmetric matrix $\iu t {\cal H}(\mathbf {k}_0) $. We refer to this transformation/evolution of the initial wave as the \textit{quantum dispersion transform}.

The probability density function associated with this wave function is given by
\begin{align}
 \rho(\mathbf {r}-(\mathbf {r}_0+\mathbf{v_g}(\mathbf {k}_0) \, t),t) & = \frac{1}{Z^2}
 | \psi_{\mathbf {k}_0}(\mathbf {r}- (\mathbf {r}_0 +\mathbf{v_g}(\mathbf {k}_0) \, t)) \ast \normalx{\mathbf {r}}{\mathbf{r}_0+\mathbf{v_g}(\mathbf {k}_0) \, t}{\iu \, t\, {\cal H}(\mathbf {k}_0) }|^2\,.
 \label{eq:time-evolution-rho-dispersion}
\end{align}
The relative entropy \eqref{eq:Relative-entropy}, which is computed by an integration over the whole space, will be independent of translations of the position $\mathbf{r}$ to $\mathbf {r}_0'=\mathbf {r}_0+\mathbf{v_g}(\mathbf {k}_0) \, t$, so to analyze the entropy we can consider a simplified density function
\begin{align}
 \rho(\mathbf {r},t)&=  \frac{1}{Z^2}| \psi_{\mathbf {k}_0}(\mathbf {r}) \ast \normalx{\mathbf {r}}{0}{\iu \, t\, {\cal H}(\mathbf {k}_0) }|^2 \, .
 \label{eq:rho-dispersion-transform-simplified}
\end{align}
 This simplified form of the dispersion transform, ignoring the translation center $\mathbf {r}_0'=\mathbf {r}_0+\mathbf{v_g}(\mathbf {k}_0) \, t$, is useful for studying the entropy of such an evolution as a quantum dispersion process.

Consider the coherent states, that is, the eigenstates of the ladder operators \eqref{eq:coherent-states} in position space, expanded to three dimensions and translated to the center position $\mathbf{r}_0$, i.e.,
\begin{align}
 \psi_{\mathbf {k}_0}(\mathbf {r}-\mathbf {r}_0)= \bra{\mathbf {r}}\ket{\alpha}=\frac{1}{Z_1}\, \normalx{\mathbf {r}}{\mathbf {r}_0}{\matrixsym{\Sigma} }\, \eu^{\iu \mathbf {k}_0\cdot \mathbf {r}}\,,
 \label{eq:coherent-state-3D}
\end{align}
where $Z_1=\int \big ( \normalx{\mathbf {r}}{\mathbf {r_0}}{\matrixsym{\Sigma} }\big)^2\diff^3 \mathbf {r}={1}/{2^3 \piu^{\frac{3}{2}}(\det \matrixsym{\Sigma})^{\frac{1}{2}}}$ is a real value normalization constant so that $\rho(\mathbf {r},0)=\psi^*_0(\mathbf {r},0)\psi_0(\mathbf {r},0)$ integrates to 1 over the entire space. The parameters $\mathbf {r_0}$, $ \matrixsym{\Sigma} $, and $ \mathbf {k_0}$ specify the center position, the spatial covariance matrix, and the center momentum, respectively.

\begin{theorem}
\label{thm:coherent-entropy}
 When a coherent state described by \eqref{eq:coherent-state-3D} evolves according to  the free particle \Schroedinger or Dirac equations, the entropy increases over time.
\end{theorem}

\begin{proof}
Applying the quantum dispersion transform \eqref{eq:time-evolution-psi-dispersion} to the initial state \eqref{eq:coherent-state-3D} we get
\begin{align}
 \psi_{\mathbf {k}_0}(\mathbf {r}-\mathbf {r}_0,t) & \approx
 \frac{\eu^{-\iu \mathbf {k}_0\cdot v_p t}}{Z_2} \normalx{\mathbf {r}}{\mathbf {r}_0 +\mathbf{v_g}(\mathbf {k}_0) t}{\matrixsym{\Sigma} } \eu^{\iu \mathbf {k}_0\cdot \mathbf {r}} \ast \normalx{\mathbf {r}}{\mathbf{r}_0+\mathbf{v_g}(\mathbf {k}_0) t}{\iu t {\cal H}(\mathbf {k}_0) ) }
 \\
 & \approx
 \frac{\eu^{\iu \mathbf {k}_0\cdot (\mathbf {r}-v_p t)}}{Z_2}\normalx{\mathbf {r}}{\mathbf {r}_0 +\mathbf{v_g}(\mathbf {k}_0) t}{\matrixsym{\Sigma}+\iu t {\cal H}(\mathbf {k}_0) }\, ,
\end{align}
which is
a wave packet moving with phase velocity $\mathbf{ v_p}\coloneqq \mathbf{\mathbf{ v_p}}(\mathbf {k}_0)$,
group velocity $\mathbf{v_g}\coloneqq \mathbf{\mathbf{ \mathbf{v_g}}}(\mathbf {k}_0)$, and with a time-varying complex value covariance $\matrixsym{\Sigma} +\iu t {\cal H}(\mathbf {k}_0)$. The probability density then becomes
\begin{align}
 \rho(\mathbf {r},t) & = \frac{1}{Z_2^2} \normalx{\mathbf {r}}{\mathbf {r}_0 +\mathbf{v_g} t}{\left (\matrixsym{\Sigma} +\iu t {\cal H} \right)} \normalx{\mathbf {r}}{\mathbf {r}_0 +\mathbf{v_g} t}{\left (\matrixsym{\Sigma} -\iu t {\cal H} \right)}
 \\
 & = \normalx{\mathbf {r}}{\mathbf {r}_0 +\mathbf{v_g} t}{\left (\left (\matrixsym{\Sigma} +\iu t {\cal H} \right)^{-1}+\left (\matrixsym{\Sigma} -\iu t {\cal H} \right)^{-1}\right )^{-1}}\,,
 \label{eq:density-coherent}
 \\
\intertext{and using \eqref{eq:sigma-t}}
 \rho(\mathbf {r},t) & = \normalx{\mathbf {r}}{\mathbf {r}_0 +\mathbf{v_g} t}{\frac{1}{2}\matrixsym{\Sigma}(t)}\,,
\end{align}
where $\matrixsym{\Sigma}(t)=\matrixsym{\Sigma} + t^2 {\cal H} \matrixsym{\Sigma}^{-1} {\cal H}$. The packet's probability density's center is moving with velocity $\mathbf{v_g}$ and its covariance $\frac{1}{2}\matrixsym{\Sigma}(t)$ is varying  over time. The entropy \eqref{eq:Relative-entropy} is then given by
\begin{align}
 \mathrm{S}_t & =-\int_{\Omega} \normalx{\mathbf {r}}{\mathbf {r}_0 +\mathbf{v_g} t}{\frac{1}{2}\matrixsym{\Sigma}(t)} \ln \normalx{\mathbf {r}}{\mathbf {r}_0 +\mathbf{v_g} t}{\frac{1}{2}\matrixsym{\Sigma}(t)}\diff^3\mathbf {r}
 \\
 & = \frac{3}{2}+\frac{3}{2}\ln\left(2\piu\right) +\frac{1}{2}\ln \det\left ( \frac{1}{2}\matrixsym{\Sigma}(t)\right)\,.
 \label{eq:entropy-coherent}
\end{align}
By Lemma~\ref{lemma:sigma-t} in Appendix~\ref{appendix:Covariance-Properties}, $\det \matrixsym{\Sigma}(t)$ increases over time.
 The logarithm function is monotonically increasing, and thus the entropy increases over time.
\end{proof}

Note however, that  we can construct an initial probability amplitude,
\begin{align}
 \psi_{\tau}(\mathbf {r}, t=0)= \frac{1}{Z} \normalx{\mathbf {r}}{\mathbf {r}_0}{\matrixsym{\Sigma} } \eu^{\iu \mathbf {k}_0\cdot \mathbf {r}} \ast \normalx{\mathbf {r}}{0}{-\iu \tau {\cal H}(\mathbf {k}_0) },
 \label{eq:coherent-state-evolved-backward-time}
\end{align}
where $\tau >0$ and such state will evolve through the quantum dispersion transform, for a period of time $\tau$, decreasing entropy. We now investigate  a partition of the Hilbert state space according to  the behavior of the time-evolution of the entropy.

\subsection{Entropy-Partition of the Hilbert Space}
\label{sec:IV}

We propose  a  novel partition of the Hilbert space  given a Hamiltonian $H$,  a time interval $\deltau t$, and an entropy $S_t$. All possible evolutions of the entropy $\mathrm{S}_t$ of a state in a time interval $[0,\deltau t]$ can be classified as belonging to one of the following four disjoint sets.
\begin{enumerate}
 \item
 ${\cal C}_{H, \deltau t}$ is the set of the evolutions for which $\mathrm{S}_t$ is a constant.
 \item
 ${\cal W}_{H, \deltau t}$ is the set of all the evolutions for which $\mathrm{S}_t$ is decreasing, but it is not a constant.
 \item
 ${\cal M}_{H, \deltau t}$ is the set of all the evolutions for which $\mathrm{S}_t$ is increasing, but it is not a constant.
 \item
 ${\cal I}_{H, \deltau t}$ is the set of all the remaining evolutions. These are oscillating evolutions in which $\mathrm{S}_t$ is strictly decreasing in some subinterval of $[0,\deltau t]$ and it is strictly increasing in another subinterval of $[0,\deltau t]$.
\end{enumerate}
To simplify the notation we drop the subscripts in the above definitions, i.e., we refer to the four sets above as $\mathcal{C}$, $\mathcal{W}$, $\mathcal{M}$, and $\mathcal{I}$.

\begin{lemma}
Set ${\cal C}$ consists of the set of all complex value wave functions of the form $\psi(\mathbf{r},t)=A(\mathbf{r}) \eu^{\iu f(\mathbf{r},t)}$ where $A(\mathbf{r})$ and $f(\mathbf{r},t)$ are real-valued functions. This set includes all stationary solutions.
\end{lemma}
\begin{proof}
 For the entropy to be constant over time, the probability density function $\rho(\mathbf{r},t)=\psi^* \psi$ must also be constant over time. The general representation of a complex-valued function is $\psi(\mathbf{r},t)=A(\mathbf{r},t) \eu^{\iu f(\mathbf{r},t)}$ with $A(\mathbf{r},t)$ and $f(\mathbf{r},t)$ being real-valued functions representing the non-negative magnitude and the phase, respectively. Thus, the probability density function becomes $\psi^* \psi=A^2(\mathbf{r},t) \eu^{-\iu f(\mathbf{r},t)} \eu^{\iu f(\mathbf{r},t)} = A^2(\mathbf{r},t)$, and for it to be constant over time, $A(\mathbf{r},t)$ must be just $A(\mathbf{r})$. The stationary solutions are of the form $\psi(\mathbf{r},t)=A(\mathbf{r}) \eu^{\iu \varphi(\mathbf{r})} \eu^{-\iu \frac{E}{\hbar} t}$ where $E$ is the energy eigenvalue of the Hamiltonian and $\phi(\mathbf{r})$ is a spatial dependent phase. Clearly, all such solutions are in ${\cal C}$ with $f(\mathbf{r},t)=\varphi(\mathbf{r})-{E}t/{\hbar} $.
\end{proof}

\begin{lemma}
Consider two stationary states
$\psi_1(\mathbf{r},t)=A_1(\mathbf{r})\eu^{\iu \varphi_{1}(\mathbf{r})} \eu^{\iu \omega_1 t}$ and $\psi_2(\mathbf{r},t)=A_2(\mathbf{r}) \eu^{\iu \varphi_{2}(\mathbf{r})} \eu^{\iu \omega_2 t}$ with $\omega_1\ne \omega_2$. If
$\deltau t > {\piu}/{|\omega_1- \omega_2|}$ then their superposition is in $\mathcal{I}$.
\end{lemma}
\begin{proof}
 The superposition is $\psi(\mathbf{r},t)= \frac{1}{Z} \left (A_1(\mathbf{r}) \eu^{\iu \varphi_{1}(\mathbf{r})} \eu^{\iu \omega_1 t}+A_2(\mathbf{r}) \eu^{\iu \varphi_{2}(\mathbf{r})} \eu^{\iu \omega_2 t}\right )$, where $Z$ is a normalization. Let $\Delta \omega = \omega_{1} - \omega_{2}$ and $\Delta \varphi (\mathbf{r})= \varphi_{1}(\mathbf{r})- \varphi_{2}(\mathbf{r})$. Then $\rho(\mathbf{r},t)=\frac{1}{Z^2} \left (A_1^2(\mathbf{r})+A_2^2(\mathbf{r})+ 2 A_1(\mathbf{r}) A_2(\mathbf{r}) \cos (\Delta \varphi(\mathbf{r}) +\Delta \omega t)\right)$. If $\deltau t > {\piu}/{|\omega_1- \omega_2|}$ then the interval $[\Delta \varphi(\mathbf{r}), \Delta \varphi(\mathbf{r}) + \Delta \omega \deltau t ]$ is longer than $\uppi$ and therefore the interval $\big(\Delta \varphi(\mathbf{r}), \Delta \varphi(\mathbf{r}) + \Delta \omega \deltau t \big)$ contains a strict local extremum of $\cos (\Delta \varphi (\mathbf{r}) +\Delta \omega t)$, which is also a strict local extremum of $\rho$. Thus, $\rho$ is oscillating and so is $\mathrm{S}_t$. Therefore the superposition is in $\mathcal{I}$.
\end{proof}

\begin{figure}
 \centering
 \includegraphics[width=0.95\linewidth]{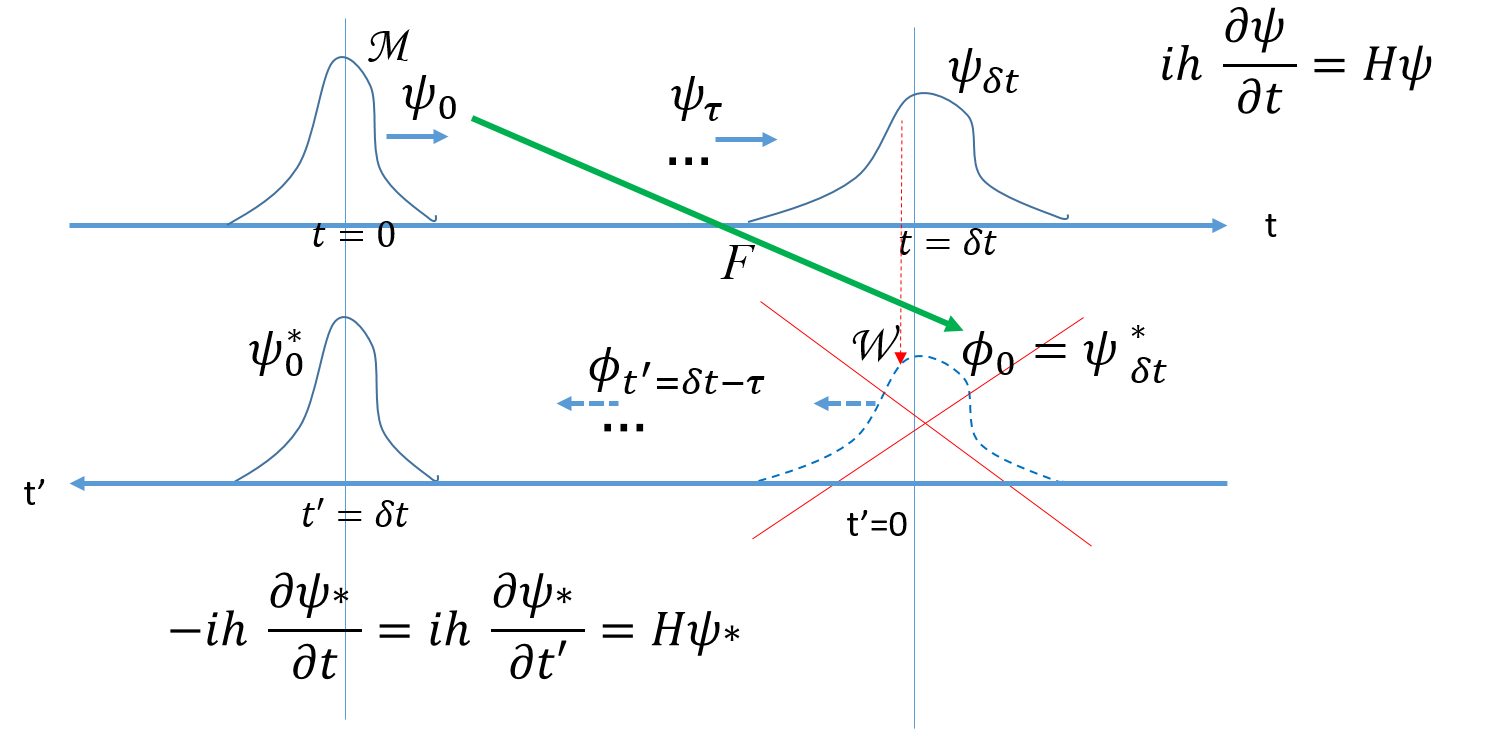}
 \caption{We depict a scenario considered in Theorem~\ref{thm:involution}. (i) Top axis t: We start at $t=0$ with a state $\psi_0 \in {\cal M}$ that evolves over a time interval $\deltau t$ according to Hamiltonian $H$, ending at state $\psi_{\deltau t}$. (ii) Bottom axis $t' =-t$: The conjugate state of $\psi_{\deltau t}$ is created as a new state $\phi_0 = \psi_{\deltau t}^*$, which is the application of $\phi_0=F(\psi_0)$ in Theorem~\ref{thm:involution}. We evolve $\phi_0$ according to $H$ for a time interval $\deltau t$, which is equivalent to evolving $\phi_0$ by the conjugate equation and backward in time, or forward in $t'$. Then, at any time $0< \tau < \deltau t$ the evolution of state $\psi_{\tau}$ will be mirroring the evolution of state $\phi_{\deltau t-\tau}$. Since $\psi_0 \in {\cal M}$, the evolution of $\phi_0$ backward in time will result in $\phi_0 \in {\cal W}$.  }
 \label{fig:Correspondence}
 \end{figure}

\begin{theorem}
 \label{thm:involution}
 Define a function $F: \ket{\psi} \mapsto \eu^{\iu H \deltau t }\ket{\psi^*}$. Then the restriction of this function to the set $\mathcal{M}$
is an involution from $\mathcal{M}$ onto $\mathcal{W}$.
\end{theorem}
\begin{proof}
Pick any $\ket{\psi}$ and let
$\ket{\psi_{t}} = \eu^{- \iu H t} \ket{\psi}$ be its time evolution. Function $F$ maps $\ket{\psi}$ to the state $\ket{\phi} = \eu^{\iu H \deltau t }\ket{\psi^*}$. Then,
 the evolution of $\ket{\phi}$ yields
$\ket{\phi_{t}} = \eu^{-\iu H t }\ket{\phi} = \eu^{\iu H (\deltau t -t)}\ket{\psi^{*}}$. Pick any $\tau \in [0,\deltau t]$. Then $\ket{\phi_{\deltau t -\tau}} = \eu^{\iu H \big(\deltau t - (\deltau t -\tau)\big)}\ket{\psi^{*}} = \eu^{\iu H \tau} \ket{\psi^{*}}=\ket{\psi^{*}_{\tau}}$. As $\ket{\phi_{\deltau t -\tau}}$ is the conjugate of $\ket{\psi_{\tau}}$, by Lemma~\ref{lemma:Entropy-conjugate} their probability density functions as well as their entropies are the same.

Thus the evolution of the entropy of $\phi_{t}$ in the time interval $[0,\deltau t]$ is the ``time-reflected image'' of the evolution of the entropy of $\psi_{t}$ in that time interval in the following sense. Let $t'=\deltau t- t$.
Then for every $t \in [0,\deltau t]$ the entropy of $\ket{\psi}$ at time $t$ is equal to the entropy of $\ket{\phi}$ at time $t'$. And as $t$ evolves forward from $0$ to $\deltau t$, $t'$ evolves backward from $\deltau t$ to $0$. Figure~\ref{fig:Correspondence} illustrates such time-reflection of the evolution of the two states. Thus the two entropies traverse the same path, but in the opposite directions. Therefore, when $\ket{\psi} \in \mathcal{M}$, it follows that $\ket{\phi} \in \mathcal{W}$.

Pick any $\ket{\phi} \in \mathcal{W}$.
Then by an argument analogous to the one used earlier in the proof, $\eu^{-\iu H \deltau t}\ket{\phi^{*}} \in \mathcal{M}$. As $F\left(\eu^{-\iu H \deltau t}\ket{\phi^{*}} \right) = \ket{\phi}$, $F$ is surjective.

We now complete the proof that $F$ is an involution. For any $\ket{\psi}$,
$F\big (F(\ket{\psi})\big) = F(e^{\iu H \deltau t}\ket{\psi^{*}}) =
e^{\iu H \deltau t}(e^{\iu H \deltau t}\ket{\psi^{*}})^{*} = e^{\iu H \deltau t} e^{-\iu H \deltau t} (\ket{\psi^{*}})^{*} =\ket{\psi}$. Thus $F^2$ is the identity function, and therefore $F$ is an involution.
\end{proof}

Moreover,

\begin{theorem}
  A single particle in $\mathcal{M}$, for every finite time interval $\delta t$,    will remain in $\mathcal{M}$ forever with the possible exception of Majorana fermions.
\end{theorem}

\begin{proof}

Theorem~\ref{thm:involution} establishes an involution from set $\mathcal{M}$ onto set $\mathcal{W}$, $F: \ket{\psi} \mapsto \eu^{\iu H  t }\ket{\psi^*}$. It is achieved  by taking a particle state  $\ket{\psi_{t}} = \eu^{- \iu H t} \ket{\psi}$   that time evolved from $\ket{\psi} \in \mathcal{M} $  to its anti-particle state  $\eu^{\iu H  t }\ket{\psi^*} \in \mathcal{W}$ via the conjugate operator. Thus, such state $\eu^{\iu H  t }\ket{\psi^*}$ must have its charges reversed with respect to the original state $\ket{\psi}$.

Particles cannot be transformed into anti-particles due to conservation laws, such as charge conservation, unless the anti-particles are the particles themselves. Majorana fermions have such a property, and it is speculated that neutrinos are Majorana particles \cite{albert2014search}. In the boson family, photons are their own anti-particles. Photons travel only at the constant speed of light and so their position entropy is constant over time and they are in $\mathcal{C} $. Thus, photons  cannot be in sets $\mathcal{M}$ or $\mathcal{W}$.  Thus, with the exception of Majorana particles, states in $\mathcal{M}$ cannot be transformed into states in $\mathcal{W}$ and vice-versa. Moreover, since the particle is in state  $\mathcal{M}$ for any finite time interval $\deltau t$, it cannot be in $\mathcal{I}$ or in $\mathcal{C}$, and  thus it will remain in   $\mathcal{M}$ forever.
\end{proof}

\section{The Entropy Law}
\label{sec:entropy-law}
In classical statistical physics, the entropy provides an arrow of time through the second law of thermodynamics. Given a probability density function in quantum physics and the entropy associated with it, is there a law analogous to the second law of thermodynamics? We have just shown that in quantum physics various scenarios already obey such a law, and thus a time arrow already exists for those scenarios.  However, there are scenarios where the current quantum theory does not guarantee that the entropy increases. Inspired by the thermodynamic law, and to ensure a time arrow, we propose  such an entropy law in quantum theory.

\begin{postulate}[\textbf{The Entropy Law}]
 \label{postulate:1}
The position entropy of a physical solution to a quantum equation is an increasing function of time.
\end{postulate}
As in the standard definitions, ``increasing,'' in contrast to ``strictly increasing,'' means ``weakly increasing.''

This law aims to treat quantum physics as an alternative statistical theory, equipped with an arrow of time via a law analogous to the second law of thermodynamics. In this way, a strictly increasing $\mathrm{S}_t$ over time in most quantum solutions implies the irreversibility of natural processes, and thus an asymmetry between the past and the future. The exception is for stationary states where $\mathrm{S}_t$ remains constant over time. Although quantum states evolve through unitary operators, for evolution in which $\mathrm{S}_t$ is increasing but is not constant, the inverse evolution must be ruled out.

This law can be tested for oscillatory solutions since according to current laws, the entropy of a single quantum particle in a oscillatory state can decrease. Also, this law can be tested for  collisions of multiple particles as we discuss next.

\subsection{Observations and Speculations}

 We speculate that particles that oscillate decay so that the entropy never decreases.  Perhaps this law could account for neutrinos oscillations. Neutrinos with specific flavors (electron, muon, and tau) are in a superposition of the three mass states. In general, they oscillate across flavors in flight \cite{superkamiondesuzuki2019} becoming the larger mass one (tau).   We wonder whether the oscillations are impacted by the entropy since one of the Hessian eigenvalues increases with mass, i.e., the dispersion is larger and so the entropy increases for larger mass and high-speed particles. Moreover, if for the entropy to increase requires that a neurtrino transform to its conjugate state, to anti-neutrino, than would justify  that neutrinos are Majorana particles as it is already speculated \cite{albert2014search}).

 We also speculate on the reason why isolated atoms are always in the ground state. For example, ``excited electrons'' in the hydrogen atom always end in the ground state, a stationary state (thus belongs to the Hilbert space set $\mathcal{C}$). We speculate that the entropy increases in such transitions (one must account for the entropy of the emitted photon). The  stationary states according to either \Schroedinger equations or Dirac equations are not stationary states under QED. But are there stationary states other than the ground state in an atom? We speculate that atoms are stable because superposition of stationary states, if ever formed, would settle to a stationary state with lower energy (ultimately the ground state), through the emission of photons and a further increase in entropy.

From the lemma it follows that for such superpositions the entropy can be increasing for at most time interval of length ${\piu}/{|\omega_1-\omega_2|}$. Therefore, according to the entropy law, such superpositions can only last for at most such time interval and then they would have to transform into new states where the entropy does not decrease. Such events would otherwise be considered spontaneous transitions, without a satisfactory cause.

We speculate that atoms are stable because superposition of stationary states, if ever formed, would settle to a stationary state with lower energy through the emission of photons and a further increase in entropy.

\subsection{The Two-Particle System}
\label{sec:two-particles}
We now consider a two-particle system evolving under Hamiltonian $H$. The particles are described by a wave packets $\psi_1(\mathbf {r}_1,t)$ and $\psi_2(\mathbf {r}_2,t)$, such that $\{ \psi_1(\mathbf {r}_1,t), \psi_2(\mathbf {r}_2,t)\} \in {\cal M}$, i.e, the entropy increases over time for each one of them separately.

Let us consider the two cases where both particles are fermions or both are bosons. The two-particle states are then described by
 \begin{align}
 \ket{\psi_t^{\fer,\bos}}&=\frac{1}{\sqrt{C_t}} \left (\ket{\psi^1_t}\ket{\psi^2_t}\mp \ket{\psi^2_t}\ket{\psi^1_t}
 \right )\,,
 \\
 \intertext{and projecting on $\bra{\mathbf {r}_1}\bra{\mathbf {r}_2}$ we get}
 \psi^{\fer,\bos}(\mathbf {r}_1,\mathbf {r}_2 ,t)&=\frac{1}{\sqrt{C_t}}
 \left ( \psi_1(\mathbf {r}_1,t) \psi_2(\mathbf {r}_2,t)\mp \psi_1(\mathbf {r}_2,t) \psi_2(\mathbf {r}_1,t) \right ),
 \label{eq:two-particle-state}
 \end{align}
 where $C_t$ is a normalization constant and the signs ``$\mp$'' represent the fermions (``$-$'') and the bosons (``+''). When the two states, $ \psi_1(\mathbf {r},t)$ and $ \psi_2(\mathbf {r},t)$, are orthogonal, we get $C_t=2$.

 The probability density function is
 \begin{align}
 \label{eq:density-two-particles}
 \rho^{\fer,\bos}(\mathbf {r}_1,\mathbf {r}_2 ,t)& =|\psi^{\fer,\bos}(\mathbf {r}_1,\mathbf {r}_2 ,t)|^2
 \\
 &=\frac{1}{C_t}\left [ \rho_1(\mathbf {r}_1,t) \rho_2(\mathbf {r}_2 ,t)+\rho_1(\mathbf {r}_2,t) \rho_2(\mathbf {r}_1 ,t) \right .
 \\
 & \quad \left . \mp
 \psi_1(\mathbf {r}_1,t) \psi_2^*(\mathbf {r}_1,t) \psi_2(\mathbf {r}_2,t)
 \psi_1^*(\mathbf {r}_2,t) \mp \psi_1(\mathbf {r}_2,t) \psi_2^*(\mathbf {r}_2,t) \psi_2(\mathbf {r}_1,t) \psi_1^*(\mathbf {r}_1,t) \right ]\, .
 \end{align}

 The relative entropy \eqref{eq:Relative-entropy},  extends to many particles as follows
\begin{align}
\mathrm{S}_t(\ket{\psi}) =-\int \diff^3\mathbf{r}_1 \int \diff^3\mathbf{r}_2 \; \cdots \; \int \diff^3\mathbf{r}_N \, \rho (\mathbf{r}_1, \hdots, \mathbf {r}_N,t) \ln \rho (\mathbf{r}_1, \hdots, \mathbf {r}_N,t) \,.
\label{eq:Relative-entropy-many-particles}
\end{align}
and so for a two particle system  it is
 \begin{align}
 \label{eq:entropy-two-particles}
 S^{\fer,\bos}\left (\ket{\psi^1_t},\ket{\psi^2_t}\right ) &= - \int \! \diff^3 \mathbf {r}_1 \int \! \diff^3 \mathbf {r}_2\,  \rho^{\fer,\bos}(\mathbf {r}_1,\mathbf {r}_2 ,t)
 \ln \rho^{\fer,\bos}(\mathbf {r}_1,\mathbf {r}_2 ,t) \, .
 \end{align}

 \subsubsection{Two Coherent Particles Moving Toward Each Other}
Let us conduct a quantitative analysis of the solution for two coherent wave packet moving towards each other
\begin{align}
\bra{\mathbf {r}}\ket{\Psi^{1}_t}=\Psi^{1}(\mathbf {r},t) &=\frac{\eu^{\iu \mathbf {k}_1\cdot (\mathbf {r}-v_p(\mathbf {k}_1) \, t)}}{Z}\, \normalx{\mathbf {r}}{\mathbf {x}_1 +\mathbf{v_g}(\mathbf {k}_1) \, t}{\matrixsym{\Sigma} +\iu \, t\, {\cal H}(\mathbf {k}_1) }\, \qquad \text{and}
\\
\bra{\mathbf {r}}\ket{\Psi^{2}_t}=\Psi^{2}(\mathbf {r},t) &=\frac{\eu^{\iu \mathbf {k}_2\cdot (\mathbf {r}-v_p(\mathbf {k}_2) \, t)}}{Z'}\, \normalx{\mathbf {r}}{\mathbf {x}_2 +\mathbf{v_g}(\mathbf {k}_2) \, t}{\matrixsym{\Sigma} +\iu \, t\, {\cal H}(\mathbf {k}_2) } \, ,
\label{eq:two-coherent-states}
\end{align}
where we already deduced by Theorem ~\ref{thm:coherent-entropy} that for each probability amplitude separately the entropy increases over time. We focus on the case where the wave packets are moving toward each other with ``complementary'' parameters, i.e.,
\begin{align}
 \mathbf{\mathbf{ v}_{\ph}}(\mathbf {k}_2)&=-\mathbf{\mathbf{ v}_{\ph}}(\mathbf {k}_1), && \text{phase velocities opposite to each other, but same magnitude; }
 \\
 \mathbf{\mathbf{ v}_{\ph}}(\mathbf {k}_1)&=v_p\, \frac{\mathbf {x}_2-\mathbf {x}_1}{|\mathbf {x}_2-\mathbf {x}_1|}, && \text{phase velocity along the segment connecting the centers;}
 \\
 \mathbf{ \mathbf{ v}_{\gr}}(\mathbf {k}_2)&=-\mathbf{ v}_{\gr}(\mathbf {k}_1), && \text{group velocities opposite to each other, but same magnitude; }
 \\
 \mathbf{ \mathbf{ v}_{\gr}}(\mathbf {k}_1)&={ v}_{\gr}\, \frac{\mathbf {x}_2-\mathbf {x}_1}{|\mathbf {x}_2-\mathbf {x}_1|}, && \text{group velocity along the path connecting the centers;}
 \\
 \mathbf {k}_2 &=-\mathbf {k}_1, && \text{zero total momentum;}
 \\
 \mathbf {k}_1& = k \frac{\mathbf {x}_2-\mathbf {x}_1}{|\mathbf {x}_2-\mathbf {x}_1|}, && \text{center momentum along the path connecting the centers;}
 \\
 \matrixsym{\Sigma}\coloneqq \matrixsym{\Sigma}_1 & =\matrixsym{\Sigma}_2, && \text{same initial covariance;}
 \\
 {\cal H}\coloneqq{\cal H}(\mathbf {k}_1) &={\cal H}(\mathbf {k}_2), && \text{same initial hessian (following from $\mathbf{k}_2=-\mathbf{k}_1$).}
\end{align}

\begin{figure}
 \centering
 \includegraphics[width=0.52\linewidth]{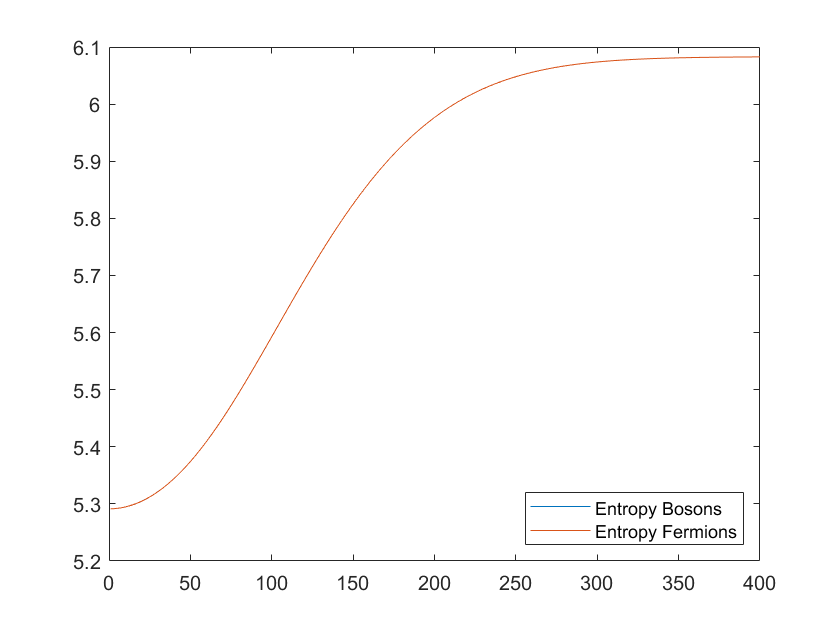}
 \hfill
 \includegraphics[width=0.45\linewidth]{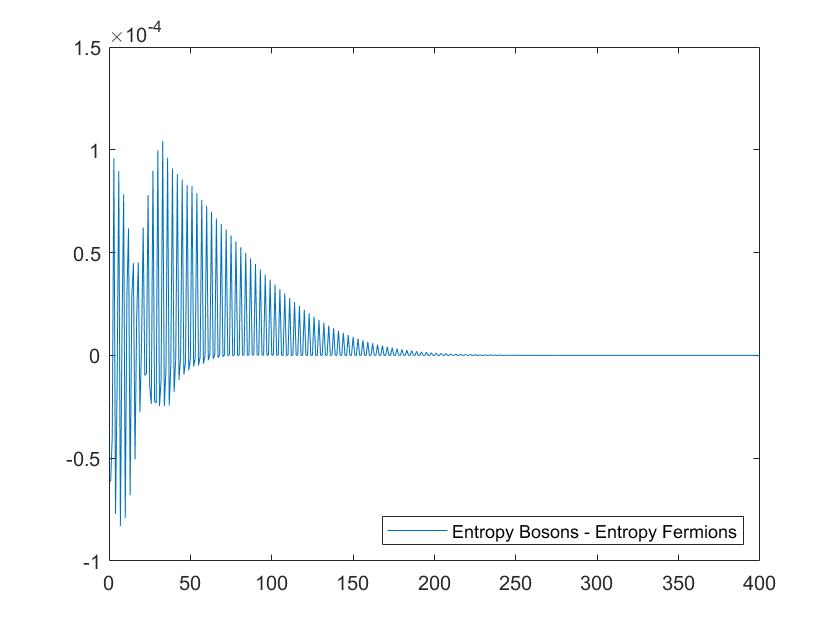}
 \\
 (a) \qquad \qquad\qquad \hfil \qquad \qquad \qquad (b)
 \caption{(a) Two one-dimensional coherent wave packets in one dimension separated by a distance $x=0,\hdots,400$\,\duu (the x-axis). The grid is discretized in $2\,000$\,\duu. For one packet $\sigma_1=50$\,\duu and for the other $\sigma_2=100$\,\duu, and the values of the center momenta are $k_{0}=1, 2$, respectively. We plot two graphs for the entropy of two wave packets, one graph describing two fermions and the other two bosons. The graphs are so similar that with one decimal order of magnitude on the entropy value, no difference can be seen at all distances $x=0,\hdots,400$\,\duu. Note that the closer the two packets are, the lower is the entropy. The interference decreases entropy. (b) We plot the smaller difference in entropy, $ \entropy(\boson) - \entropy(\fermion)$, as the distance increases. Despite being very small (order of magnitude $\approx e^{-5}$) there is a region when they are apart, $\approx [50,100]$\,\duu, where the entropy of a boson pair is consistently larger). As they get even closer, $\approx [0,50]$\,\duu the entropy oscillates for both pairs. }
 \label{fig:two-particles}
 \end{figure}

Combining  \eqref{eq:two-coherent-states} to form the probability amplitude \eqref{eq:two-particle-state},   yields the density \eqref{eq:density-two-particles} for the two coherent state.
In this our plots, \duu and \tuu  indicate synthetic spatial and temporal units, respectively,  employed in the simulation.
We first study this density  as a function of the distance between the particles to learn potential impact of the interference on the entropy. We plot the entropy \eqref{eq:entropy-two-particles} as a function of the distance between the two particles (see Figure~\ref{fig:two-particles}a.), and we see that the closer the particles are, the larger the  interference, and the more the entropy decreases. Both scenarios, two fermions or two bosons, yield very similar entropy values, so in Figure~\ref{fig:two-particles}a. the graphs for fermions and bosons are the same in the first orders of magnitude and cannot be distinguished. We then show in Figure~\ref{fig:two-particles}b. the difference in the entropy between the two bosons and the two fermions. In the region of large interference, the  bosons entropy is slightly larger. When the particles are very close to each other, a small oscillation occurs where either one of them can have a slightly larger entropy.

\begin{figure}
 \centering
 \includegraphics[width=0.49\linewidth]{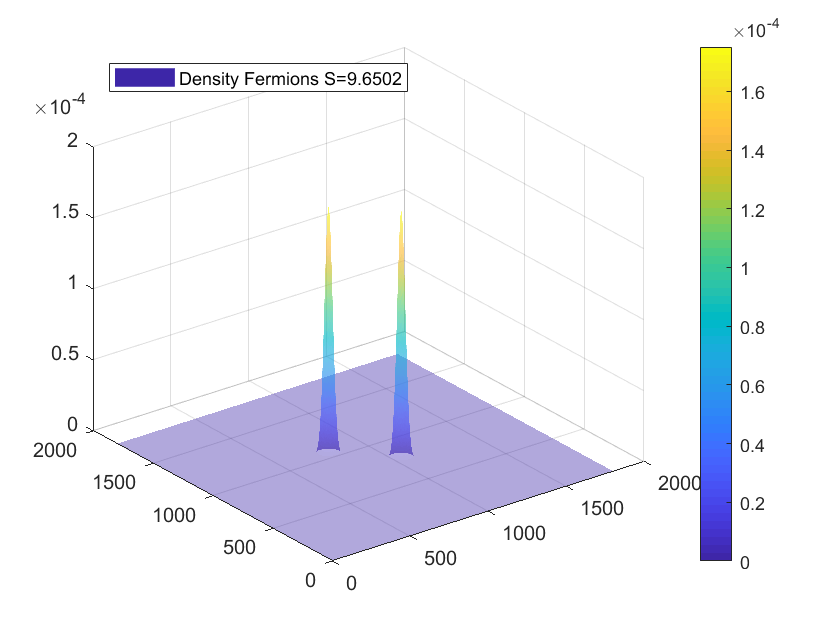}
 \hfil
 \includegraphics[width=0.49\linewidth]{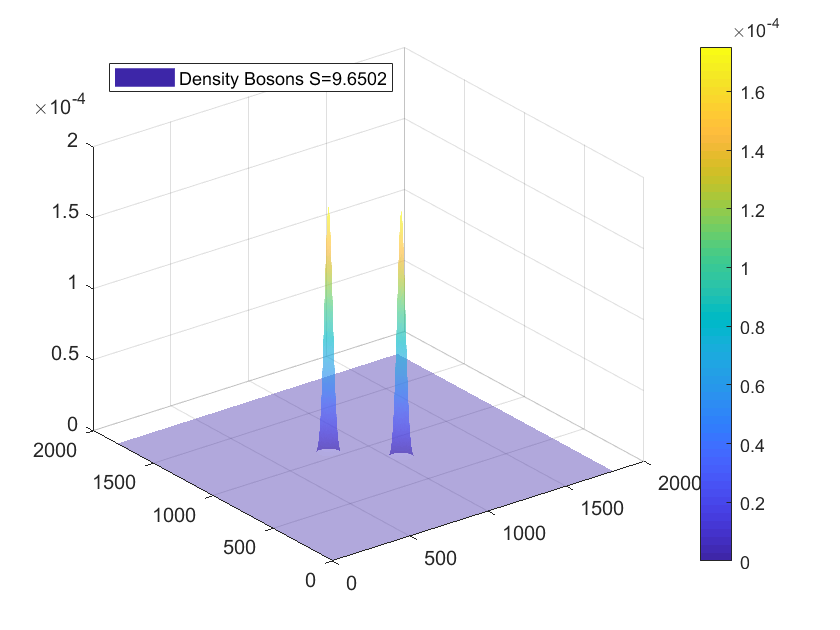}
 \\
 (a) $t=10$\,\tuu, fermions. \hfil (b) $t=10$\,\tuu, bosons.
 \\
 \centering
 \includegraphics[width=0.49\linewidth]{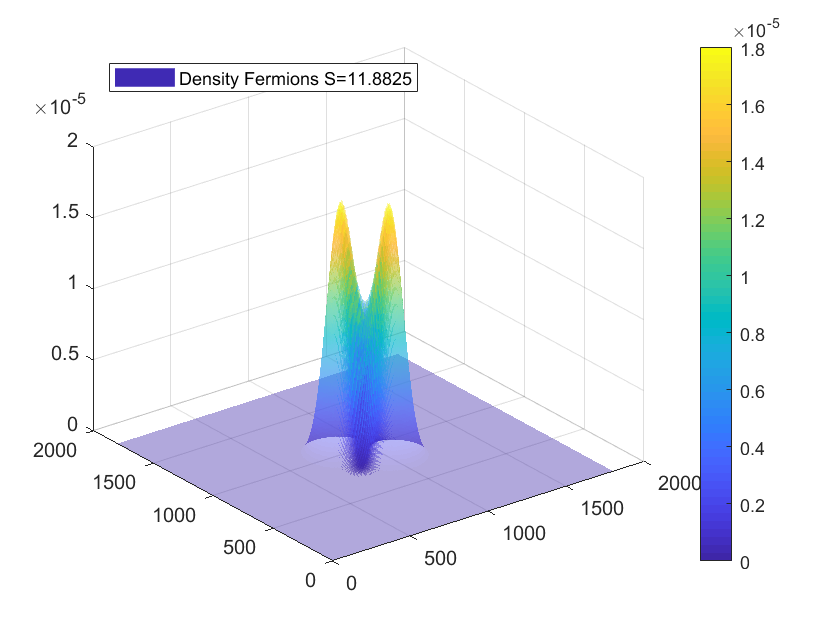}
 \hfil
 \includegraphics[width=0.49\linewidth]{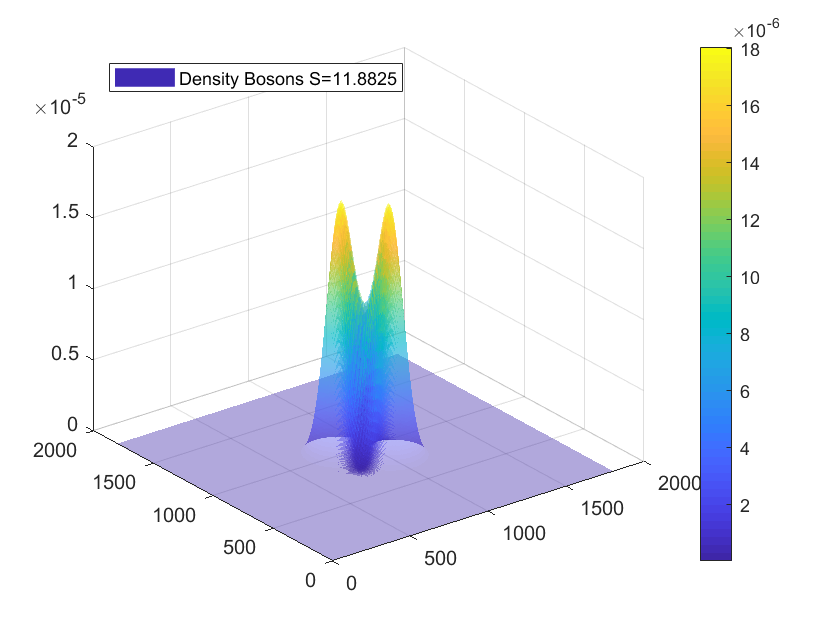}
 \\
 (c) $t=30$\,\tuu, fermions. \hfil (d) $t=30$\,\tuu, bosons.
 \\
 \centering
 \includegraphics[width=0.49\linewidth]{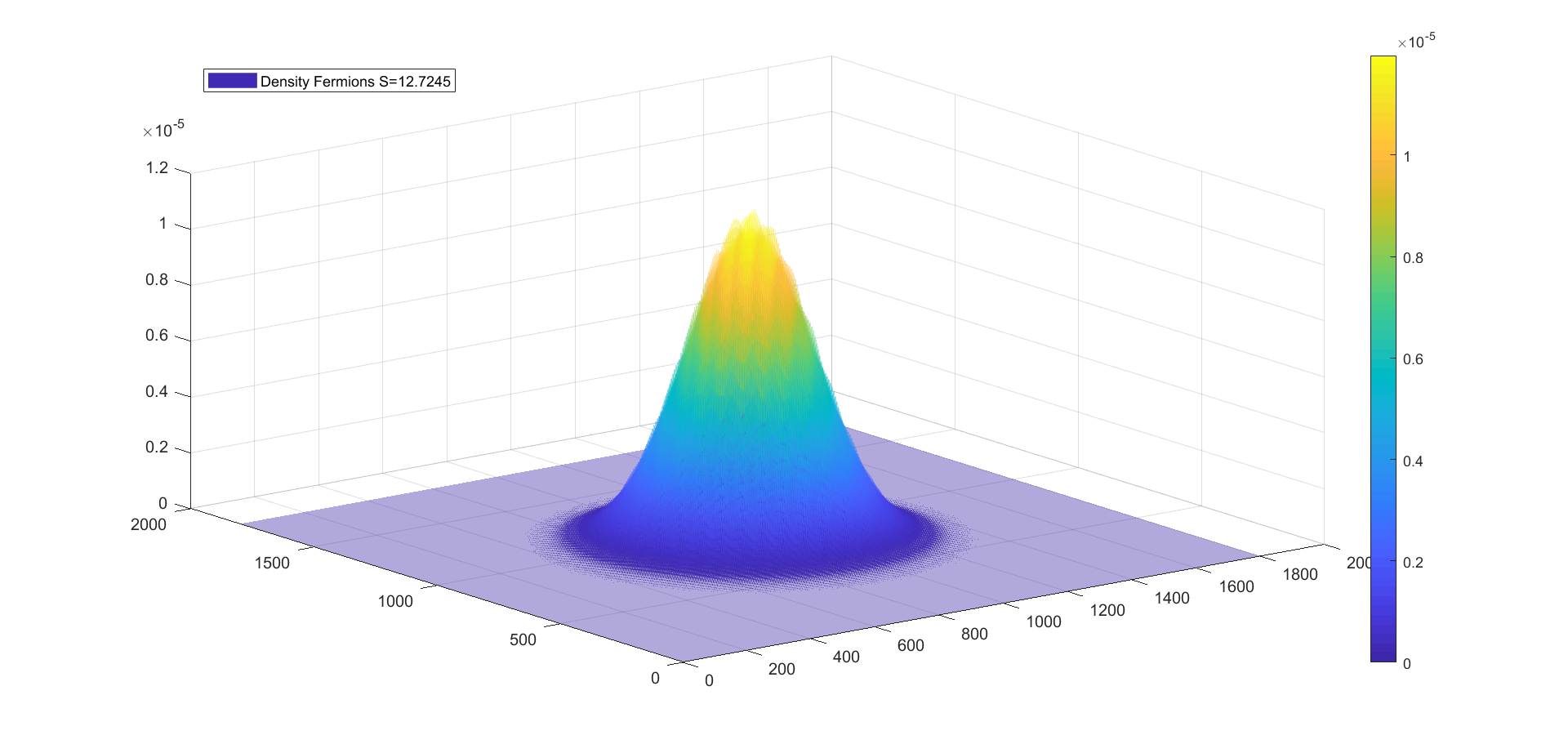}
 \hfil
 \includegraphics[width=0.49\linewidth]{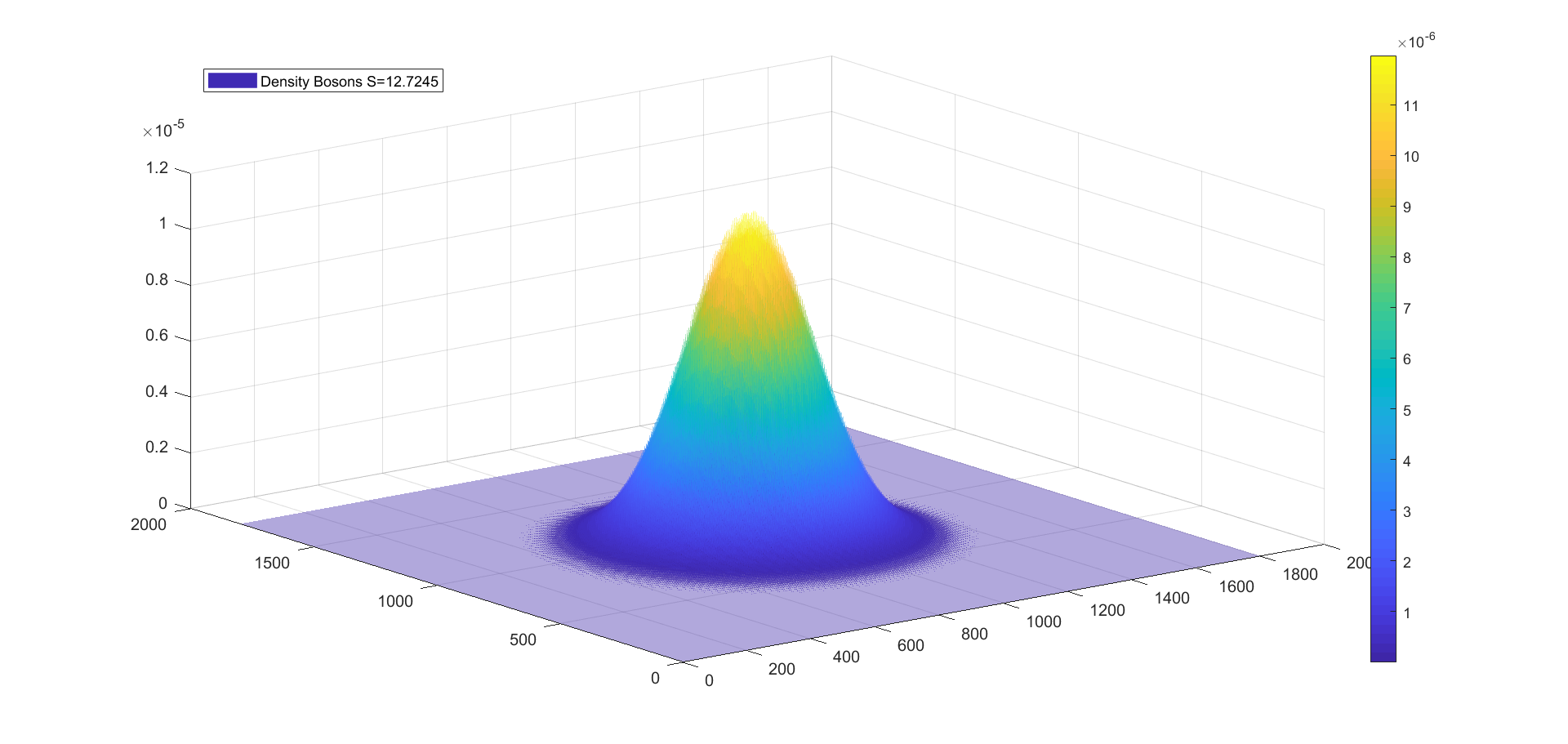}
 \\
 (e) $t=70$\,\tuu, fermions. \hfil (f) $t=70$\,\tuu, bosons.
 \caption{Two one-dimensional coherent wave packets moving towards each other with same speed (in opposite directions). Parameters: 2D grid-xy $[0,1\,800] \times [0,1\,800]$ distance units square ($\duu^2$), where the density $\rho(x,y)$ is defined (see \eqref{eq:density-two-particles}). Time $t=0,1, ...,70$\.\tuu. Speed $v_{g}=2$\,$\duu/\tuu$, momentum $k_{1}=2\piu$\,$\duu^{-1}$, standard deviation $\sigma=3$\,\duu, ${\cal H}=10$\,$\duu^{-2}$ units, initial position $x_0=750$\,\duu, $y_0=1050$\,\duu, yielding $75$\,\tuu to collide at $(900,900)$.}
 \label{fig:two-particles-evolution-states}
 \end{figure}

\begin{figure}
 \centering
 \includegraphics[width=0.49\linewidth]{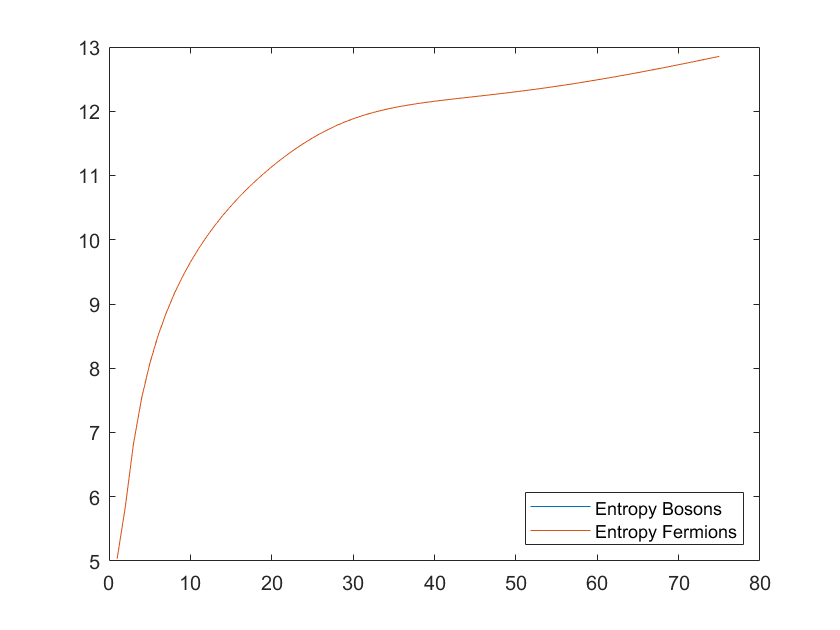}
 \hfil
 \includegraphics[width=0.49\linewidth]{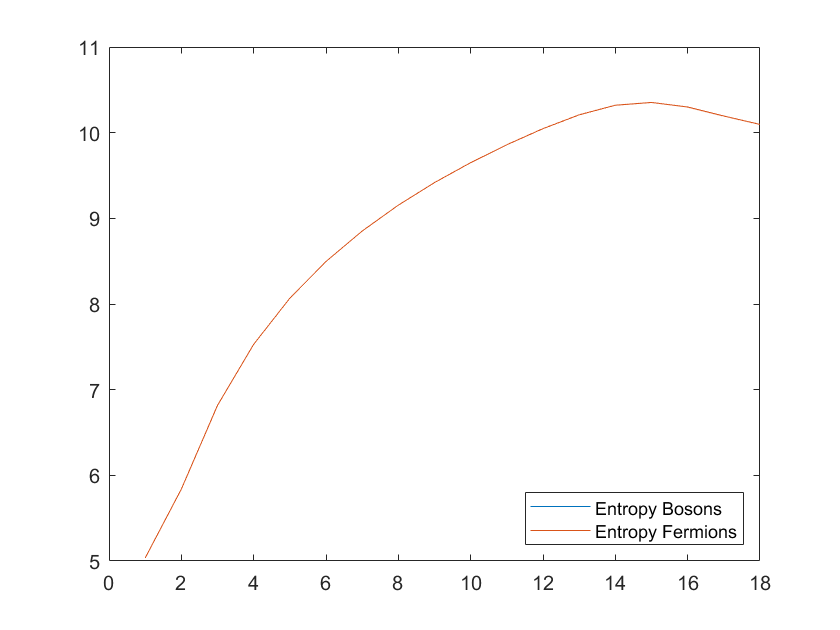}
 \\
 (a)\qquad \qquad \hfil \qquad \qquad (b)
 \caption{Two one-dimensional coherent wave packets moving towards each other with the same speed (in opposite directions). Parameters: 2D grid-xy where the density $\rho(x,y)$ (see \eqref{eq:density-two-particles}) $[0,1\, 800] \times [0,1\, 800]$ distance units square ($\duu^2$). Time $t=0,1, ...,70$\,\tuu. Momentum $k_{1}=2\piu$\,$\duu^{-1}$, standard deviation $\sigma=3$\,\duu, ${\cal H}=10$\,$\duu^{-2}$, initial position $x_0=750$\,\duu, $y_0=1050$\,\duu, on a grid of $(x,y) \in [-900,900] \times [-900,900] $, yielding $75$\,\tuu to collide at $(0,0)$. Entropy per $1$\,\tuu, when (a) speed $\abs{\mathbf{v}_\mathrm{g}}=2$\,$\duu/\tuu$ and (b) speed $\abs{\mathbf{v}_\mathrm{g}}=8$\,$\duu/\tuu$. }
 \label{fig:two-particles-evolution-entropy}
 \end{figure}

 In the next simulation we study a time evolution of the two coherent state. We set the temporal position parameters
 \begin{align}
 \mathbf {x}_1 (t)&=\mathbf {x}_1 +\mathbf{ v}_{\gr}(\mathbf {k}_1) t\,, \qquad \text{and}
 \\
 \mathbf {x}_2 (t)&=\mathbf {x}_2 -\mathbf{ v}_{\gr}(\mathbf {k}_1) t\,,
 \end{align}
 and simulated a collision of two particles, with the results plotted in Figure~\ref{fig:two-particles-evolution-states}.  The entropy of each individual particle increases due to dispersion. However, due to the interference of the two particles, the entropy of the system may decrease, as seen in Figure~\ref{fig:two-particles}. These two conflicting effects working in tandem produce the following behaviors.
As seen in Figure~\ref{fig:two-particles-evolution-entropy}a.,
in slow collisions, the entropy increases as the dispersion dominates.
And as seen in
Figure~\ref{fig:two-particles-evolution-entropy}b.,
in fast collisions, when the particles come close to each other, the entropy decreases as the interference dominates.
 Qualitatively, this behavior  suggests that as particles collide at fast speeds and before the entropy could start decreasing, the particles transform into new particles so that entropy increases while respecting the conservation laws. This discussion is qualitative, and a more experimental driven analysis taking into account real value parameters is needed.

\section{Conclusions}
\label{sec:conclusion}

Classical physics laws are time reversible and a time arrow appears in physics only when statistics of multiple particles are introduced. Such statistics are outside the physics laws governing an individual particle. Quantum theory is equipped with a position probability density $\rho(\mathbf {r},t)$, which evolves over time. The entropy associated with the probability density measures the localization information of $\rho(\mathbf {r},t)$.  We postulated a law that the location information as measured by the entropy, increases over time. The entropy function we considered is given by \eqref{eq:Relative-entropy}. One could consider an entropy function that is an extension of Shannon entropy to continuous domains as outlined by Jaynes \cite{jaynes1965gibbs} or still other functions. Our insight into why physical laws cause an information loss over time is the dispersion property of the free particle's \Schroedinger Hamiltonian and its Dirac Hamiltonian. We showed that coherent states probability density functions describing wave packets would disperse over time.

We then examined a partition of the Hilbert space induced by an entropy measure and a time interval into four sets, namely, $\mathcal{ M}$ (increasing entropy but not constant), $\mathcal{ W}$ (decreasing entropy but not constant), $\mathcal{ C}$ (entropy constant), and $\mathcal{ I}$ (entropy oscillating).

An involution between sets $\mathcal{M}$ and $\mathcal{ W}$ was established, which did help to understand the role of the conjugation process to the time arrow. We showed that assuming a quantum motion equation, and application of the conjugate operator to a state in $\mathcal{M}$ that evolves for the given time interval, will result in a state in $\mathcal{W}$. Thus, the entropy law does not allow such a conjugate process for states in $\mathcal{M}$. Still, according to the entropy law, the same conjugation process would happen for states in $\mathcal{I}$ at the time that an entropy decrease ``kicks in.'' We speculate that free neutrinos, perhaps in a superposition of states, are in states in $\mathcal{I}$ and can exist only during that part of the evolution when the entropy increases. According to the entropy law, particles evolving according to a state in $\mathcal{I}$ must have no charge so that the conjugate process does not violate the conservation law of charges. Thus, for particles without a charge, states in the set $\mathcal{I}$ will be allowed only for small time intervals. The entropy law disallows the entire set $\mathcal{W} $.

In light of the results above, we reviewed the role of the conjugate operator in the understanding of anti-particles. We proposed a similar statememt to the Feynman-Stueckelberg: \emph{There is an equivalence between describing a particle by the probability amplitude and its motion equation, and describing a particle by its conjugate and the evolution by the adjoint of the motion equation.} The default choice of representation is the one that has positive energy with the time parameter going forward.

We studied and performed some simple simulations for the collision of two fermions as well as two bosons. As the particles evolve, the entropy of each probability amplitude alone increases, but as they come close to each other, and interference occurs, an effect of decreasing the entropy occurs. These two effects compete for the entropy's behavior during the evolution. We showed in our simulations that for slow-speed collisions, the entropy might increase over time and for fast-speed collision, the entropy will start to derease at some close distance when the interference effect dominates. In these cases, according to the entropy law, a transformation into two photons moving away from each other occurs. We observed that while the entropy during the collision of two bosons and two fermions are very similar, the interference effect differs and the entropy differs at close distances.

We speculate that some physical phenomena, such as high-speed collision $e^+ + e^- \rightarrow 2 \gamma$, produce new particles so that the entropy increases (while respecting conservation laws).

In summary, the entropy law provides additional constraints on physical scenarios, beyond those provided by conservation laws.

\paragraph{Acknowledgments:} This material is based upon work supported by both the National Science Foundation under Grant No. DMS-1439786 and the Simons Foundation Institute Grant Award ID 507536 while the first author was in residence at the Institute for Computational and Experimental Research in Mathematics in Providence, RI, during the spring 2019 ``Computer Vision'' semester program.

\pagebreak
\appendix

\section{\Schroedinger Equation in the Energy-Momentum Space}
\label{eq:Schroedinger-momentum}
A one-particle system is described by the state $\ket{\Psi_t}$ that evolves over time. For non-relativistic fermions, the \Schroedinger equation describes the time evolution of any state as

\begin{align}
\iu \hbar {\frac {\partial }{\partial t}}\Psi (\mathbf {r} ,t)=\left[{\frac {-\hbar ^{2}}{2m }}\nabla ^{2}+V(\mathbf {r})\right] \Psi (\mathbf {r} ,t)\,,
\label{eq:Schroedinger-appendix}
\end{align}
where $m$ is the particle's mass, V is its potential energy, $\nabla ^{2}$ is the Laplacian (a differential operator), and $\Psi(\mathbf {r} ,t)=\bra{\mathbf {r}}\ket{\Psi_t}$ is the wave function (the state described in the coordinate bases).

The wave function $\Psi (\mathbf {r} ,t)$ can be written in the inverse Fourier basis with Minkowsky metric $\ket{-,+,+,+}$ as
\begin{align}
\Psi (\mathbf {r} ,t)&={\frac {1}{({\sqrt {2\piu }})^{3}}}\int \diff \omega \int \diff^{3}\mathbf {k} \, \Phi (\omega, \mathbf {k} )\, \eu^{-\iu(\omega t-\mathbf {k} \cdot \mathbf {r} )} \, ,
\label{eq:fourier-transform}
\end{align}
 where $\Phi (\omega, \mathbf {k})=\bra{\omega,\mathbf {k}}\ket{\Psi}$ is the Fourier transform of $\Psi (\mathbf {r} ,t)$. The \Schroedinger equation in the energy-momentum space is given by
\begin{align}
 \hbar \, \omega \, \Phi (\omega, \mathbf {k})& ={\frac {\hbar ^{2} \mathbf {k}^2}{2m }} \Phi (\omega, \mathbf {k}) + \int \diff^{3}\mathbf {k}' \, \tilde{V}( \mathbf {k}-\mathbf {k}') \Phi (\omega, \mathbf {k}')\,,
 \\
\intertext{implying}
 \omega&={\frac{\hbar \mathbf {k}^2}{2m } +\frac{1}{\hbar \, \Phi (\omega, \mathbf {k})} \tilde{V}( \mathbf {k})\ast \Phi (\omega, \mathbf {k})}
\label{eq:Schroedinger-Fourier-energy}\,,
\end{align}
where $\tilde{V}( \mathbf {k})$ is the Fourier transform of $V(\mathbf {r})$. This second term is the convolution in $\mathbf {k}$. For a free particle with no potential energy, we get \eqref{eq:Fourier-Hamiltonians}.

 \section{Dirac Equation in the Energy-Momentum Space}
 \label{sec:Dirac-momentum}
 The Dirac equation is the relativistic equation for fermions. It is described by
 \begin{align}
 (\iu \hbar \gamma ^{\mu }\partial_{\mu } -mc)\Psi =0 \quad {\rm or} \quad (\iu \hbar \slashed{\partial}_{\mu } -mc)\Psi =0\,,
 \label{eq:Dirac-standard}
 \end{align}
 where $\mu=0,1,2,3$ describe the time index and the three spatial indices with metric $\ket{-,+,+,+}$, $\gamma ^{\mu }$ are the $4 \times 4$ matrices satisfying the Clifford algebra, and in the Weyl (chiral) basis they are
 \begin{equation}
   \label{eq:1}
   \matrixsym{\gamma}^{0} = \begin{bmatrix}
 0& I_2 \\ I_2 & 0
 \end{bmatrix} \quad \text{and} \quad \matrixsym{\gamma}^{i} = \begin{bmatrix}
 0 & \sigma^i \\ -\sigma^i & 0
 \end{bmatrix}, \quad \text{for} \quad  i=1,2,3 \,,
 \end{equation}
with the Pauli matrices
 \begin{align}
 \sigma_{1}=\sigma_{x}=\begin{pmatrix}0&1\\1&0\end{pmatrix}, \qquad \sigma_{2}=\sigma _{y}&=\begin{pmatrix}0&-\iu \\ \iu & 0\end{pmatrix}, \quad \text{and} \quad \sigma _{3}=\sigma _{z}=\begin{pmatrix}1&0\\0&-1\end{pmatrix}\,.
 \end{align}
 The Dirac equation for the adjoint wave $\overline{\Psi}= \Psi^{\dag} \gamma^0 $ is
 \begin{align}
 \overline{\Psi} (\iu \hbar \gamma^{\mu } \partial _{\mu } +mc) =0 \quad {\rm or} \quad \overline{\Psi} (\iu \hbar \slashed{\partial} _{\mu } + mc)=0\,.
 \label{eq:Dirac-adjoint}
 \end{align}
 It is clear that the adjoint wave function satisfies the same equation as the wave function but with the momentum and the energy signs reversed.

 The Dirac equation \eqref{eq:Dirac-standard} can be written, similarly to \eqref{eq:Schroedinger}, as
 \begin{align}
 \iu \hbar \frac {\partial }{\partial t}
 \Psi(\mathbf {r} ,t)
 =\matrixsym{H}(\mathbf {p})\,
 \Psi(\mathbf {r} ,t)
\quad {\rm and} \quad
-\iu \hbar \frac {\partial }{\partial t}
 \overline \Psi(\mathbf {r} ,t)
 =\overline \Psi(\mathbf {r} ,t)\matrixsym{H}(-\mathbf {p})\,,
 \label{eq:Dirac-energy}
 \end{align}
 where the wave function $\Psi $ is a bispinor with the Hamiltonian
 \begin{align}
 \matrixsym{H}(\mathbf {p})= \gamma^0 \left (\matrixsym{\gamma} \cdot\matrixsym{p} c+ m c^2\right)=\gamma^0 \left (- \iu \hbar c \matrixsym{\gamma} \cdot \nabla + m c^2 \right )\, .
 \label{eq:Hamiltonian-Dirac-II}
 \end{align}

 In term of the Fourier description, position-time becomes energy-momentum. More precisely,
 \begin{align}
 \Psi(\mathbf {r} ,t) &= \frac {1}{({\sqrt {2\piu }})^{3}} \int \diff \omega \int \diff^3\mathbf {k}\, \Phi (\omega , \mathbf {k})
\eu^{-\iu \left (\omega t-\mathbf {k}\cdot \mathbf {r}\right )}
\\
 \overline \Psi(\mathbf {r} ,t) &=\Psi^{\dag} (\mathbf {r}) \gamma^0=\frac {1}{({\sqrt {2\piu }})^{3}} \int \diff \omega \int \diff^3\mathbf {k}\, \Phi^{\dag} (\omega , \mathbf {k}) \gamma^0
\eu^{\iu \left (\omega t-\mathbf {k}\cdot \mathbf {r}\right )}
\label{eq:Dirac-Fourier-timespace}
 \end{align}
 where the energy-momentum $(E,\mathbf {p})$ are described by the variables $(\omega = {E}/{\hbar}, \mathbf {k}={\mathbf {p}}/{\hbar})$. Applying these expressions to \eqref{eq:Dirac-energy} we obtain the Dirac equation and its adjoint in Fourier space
 \begin{align}
 \omega \, \Phi (\omega , \mathbf {k}) & = \gamma^0 (c \matrixsym{\gamma}\cdot \matrixsym{k} + \frac{m}{\hbar} c^2)\, \Phi (\omega , \mathbf {k}) \qquad \text{and}
 \\
 \omega \, \overline{\Phi}(\omega , \mathbf {k}) & = \overline{\Phi} (\omega , \mathbf {k}) \gamma^0 (c \matrixsym{\gamma}\cdot \matrixsym{k} + \frac{m}{\hbar} c^2)\,.
 \label{eq:dirac-momemntum-space-0}
 \end{align}
 These two equations in matrix form can be written  as an eigenvalue/eigenvector pair of equations
\begin{align}
\omega\, \gamma^0 \Phi (\omega , \mathbf {k}) & =
\begin{pmatrix}
c (\matrixsym{\sigma}\cdot \matrixsym{k} ) & \frac{m}{\hbar} c^2
\\[1.5ex]
\frac{m}{\hbar} c^2 & -c(\matrixsym{\sigma}\cdot \matrixsym{k})
\end{pmatrix} \, \gamma^0 \Phi (\omega , \mathbf {k})
\label{eq:Dirac-Momemtum-Matrix}\qquad \text{and}
 \\[1.5ex]
\omega\, \overline{\Phi}(\omega , \mathbf {k}) &= \overline{\Phi} (\omega , \mathbf {k}) \begin{pmatrix}
c (\matrixsym{\sigma}\cdot \matrixsym{k} ) & -\frac{m}{\hbar} c^2
\\[1.5ex]
-\frac{m}{\hbar} c^2 & -c(\matrixsym{\sigma}\cdot \matrixsym{k})
\end{pmatrix}\,,
 \label{eq:Dirac-Momemtum-Adjoint-Matrix}
 \end{align}
where $\matrixsym{\sigma}\cdot \matrixsym{k}=\begin{pmatrix}
 k_3 & k_1-\iu k_2 \\ k_1+\iu k_2 & -k_3
 \end{pmatrix} $, the eigenvalues are $\omega$, and the eigenvectors are $\gamma^0 \Phi (\omega , \mathbf {k}) $ or $\overline{\Phi} (\omega , \mathbf {k}) $.

The determinant of $\begin{pmatrix}
c (\matrixsym{\sigma}\cdot \matrixsym{k} ) & \frac{m}{\hbar} c^2
\\
\frac{m}{\hbar} c^2 & -c(\matrixsym{\sigma}\cdot \matrixsym{k})
\end{pmatrix} $ is $
\left ( (\frac{m}{\hbar} c^2)^2 +c^2 \matrixsym{k}^2\right )^2$.
The eigenvalues are easily derived to be $\omega(\matrixsym{k}^2)=\pm c \sqrt{ \matrixsym{k}^2+\frac{m^2}{\hbar^2} c^2}$, each appearing in multiples of two.

\section{Covariance Properties of the Time Evolution of Coherent States}
\label{appendix:Covariance-Properties}

In order to obtain an expression for the covariance of a coherent state, we start from \eqref{eq:density-coherent} and prove the following Lemma.

 \begin{lemma}
 Let
\begin{align}
 \matrixsym{\Sigma}(t)&= 2\, \left [
 (\matrixsym{\Sigma} +\iu \, t\, \mathcal{H})^{-1}+(\matrixsym{\Sigma} -\iu \, t\, \mathcal{H})^{-1}\right]^{-1}\, ,
 \end{align}
 where $ \matrixsym{\Sigma}=\matrixsym{\Sigma}(0)$. Then
\begin{align}
 \matrixsym{\Sigma}(t)&=\matrixsym{\Sigma} + t^2\, \mathcal{H} \matrixsym{\Sigma}^{-1} \mathcal{H}
 \label{eq:sigma-t}
\end{align}
\label{lemma:sigma-t-proof}
 \end{lemma}
 \begin{proof}
We start with
\begin{align}
 (\matrixsym{\Sigma} +\iu t \mathcal{H})^{-1}&= (\matrixsym{\Sigma} +\iu t \mathcal{H})^{-1} (\matrixsym{\Sigma} -\iu t \matrixsym{\Sigma} \mathcal{H} \matrixsym{\Sigma}^{-1})^{-1} (\matrixsym{\Sigma} -\iu t \matrixsym{\Sigma} \mathcal{H} \matrixsym{\Sigma}^{-1})
 \\
 &= (\matrixsym{\Sigma}^2 + t^2 \matrixsym{\Sigma} \mathcal{H} \matrixsym{\Sigma}^{-1} \mathcal{H})^{-1}(\matrixsym{\Sigma} -\iu t \matrixsym{\Sigma} \mathcal{H} \matrixsym{\Sigma}^{-1})
 \\
 &= (\matrixsym{\Sigma} + t^2 \mathcal{H} \matrixsym{\Sigma}^{-1} \mathcal{H})^{-1} -\iu t (\matrixsym{\Sigma} \mathcal{H}^{-1} \matrixsym{\Sigma}+ t^2 \mathcal{H})^{-1}\,.
\end{align}
 Similarly, $(\matrixsym{\Sigma} -\iu t \mathcal{H})^{-1}=(\matrixsym{\Sigma} + t^2 \mathcal{H} \matrixsym{\Sigma}^{-1} \mathcal{H})^{-1}+\iu t (\matrixsym{\Sigma} \mathcal{H}^{-1} \matrixsym{\Sigma}+ t^2 \mathcal{H})^{-1}$.
 Adding the two, we get
 \begin{align}
 (\matrixsym{\Sigma} +\iu t \mathcal{H})^{-1}+(\matrixsym{\Sigma} -\iu t \mathcal{H})^{-1}& =2\left(\matrixsym{\Sigma} + t^2 \mathcal{H} \matrixsym{\Sigma}^{-1} \mathcal{H}\right)^{-1}=\left (\frac{1}{2}
 \matrixsym{\Sigma}(t) \right)^{-1}
 \end{align}
 \end{proof}

 \begin{lemma}
 $\det \matrixsym{\Sigma}(t) $ increases over time.
 \label{lemma:sigma-t}
 \end{lemma}
 \begin{proof}
 $\matrixsym{\Sigma}$ is a covariance matrix with all positive eigenvalues, and thus, for any vector $\ket{v}$, $\expval{\Sigma}{v} > 0$ and we can write $\matrixsym{\Sigma}=\matrixsym{A}\matrixsym{A}^{\tran} $. Then, $\matrixsym{\Sigma}^{-1}=\matrixsym{B}\matrixsym{B}^{\tran} $, where $\matrixsym{B}^{\tran}=\matrixsym{A}^{-1}$. The real-valued matrix $\mathcal{H}$ with all positive eigenvalues is also positive definite. Using  $\mathcal{H}=\mathcal{H}^{\tran}$, we can write $\mathcal{H} \matrixsym{\Sigma}^{-1} \mathcal{H}=\mathcal{H} \matrixsym{B}\matrixsym{B}^{\tran}\mathcal{H}^{\tran} = \matrixsym{C}\matrixsym{C}^{\tran}$, where $\matrixsym{C}=\mathcal{H}\matrixsym{B}$. Thus, $\mathcal{H} \matrixsym{\Sigma}^{-1} \mathcal{H}$ is also a covariance matrix and for any vector $\ket{v}$, $\expval{\mathcal{H}\matrixsym{\Sigma}^{-1}\mathcal{H}}{v} > 0$. Thus, for any vector $\ket{v}$, $\expval{\mathcal{H}\matrixsym{\Sigma}^{-1}\mathcal{H}}{v} = \expval{\mathcal{H} \matrixsym{B}\matrixsym{B}^{\tran}\mathcal{H}^{\tran}}{v}=\expval{ \matrixsym{B}\matrixsym{B}^{\tran}}{u} =\expval{\matrixsym{\Sigma}^{-1}}{u}> 0$, where $u=\mathcal{H}^{\tran} v$. Thus, for any vector $\ket{v}$, $\expval{\matrixsym{\Sigma}(t)}{v}=\expval{\matrixsym{\Sigma}}{v}+t^2 \expval{H\matrixsym{\Sigma}^{-1}H}{v}> 0$ for $ t\in[0,\infty)$. Also, $\frac{\partial}{\partial t}\expval{\matrixsym{\Sigma}(t)}{v}=2t \expval{H\matrixsym{\Sigma}^{-1}H}{v} \ge 0 $ for $ t\in [0,\infty)$ with equality only at $t=0$. Thus, all the eigenvalues of $ \matrixsym{\Sigma}(t) $ are positive and increase over time. Therefore, $\det \matrixsym{\Sigma}(t) $ also increases over time.
 \end{proof}

 \bibliographystyle{abbrv}
\bibliography{gk01}
\end{document}